\documentclass[twocolumn]{aastex631}

\defcitealias{Guilera22b}{Guilera et al. 2023}
\usepackage{amsmath}

\usepackage{pdfrender}
\usepackage[utf8]{inputenc}
\usepackage{graphicx,color}
\usepackage{ulem,soul}

\usepackage{diagbox}

\usepackage{parskip}
\shorttitle{Quantifying the Impact of the Dust Torque on the Migration of Low-mass Planets}
\shortauthors{Guilera, Benitez-Llambay, Miller Bertolami \& Pessah}

\begin{document}

\title{Quantifying the Impact of the Dust Torque on the Migration of Low-mass Planets}

\correspondingauthor{Octavio M. Guilera}
\email{oguilera@fcaglp.unlp.edu.ar}

\author[0000-0001-8577-9532]{Octavio M. Guilera}
\affiliation{Instituto de Astrof\'{\i}sica de La Plata (IALP), CCT La Plata-CONICET-UNLP, Paseo del Bosque s/n, La Plata, Argentina.}
\affiliation{N\'ucleo Milenio de Formaci\'on Planetaria (NPF), Chile.}
\author[0000-0002-3728-3329
]{Pablo Benitez-Llambay}
\affiliation{Facultad de Ingenier\'ia y Ciencias, Universidad Adolfo Ib\'a\~nez, Av. Diagonal las Torres 2640, Pe\~nalol\'en, Chile.}
\affiliation{Data Observatory Foundation, ANID Technology Center No. DO210001}
\author[0000-0001-8031-1957
]{Marcelo M. Miller Bertolami}
\affiliation{Instituto de Astrof\'{\i}sica de La Plata (IALP), CCT La Plata-CONICET-UNLP, Paseo del Bosque s/n, La Plata, Argentina.}
\author[0000-0001-8716-3563
]{Martin E. Pessah}
\affiliation{Niels Bohr International Academy, Niels Bohr Institute, Blegdamsvej 17, DK-2100 Copenhagen Ø, Denmark}
\affiliation{School of Natural Sciences, Institute for Advanced Study, 1 Einstein Drive, Princeton, NJ 08540, USA}




\begin{abstract}
Disk solids are critical in many planet formation processes, however, their effect on planet migration remains largely unexplored. Here we assess for the first time this important issue by building on the systematic measurements of dust torques on an embedded planet by \citet{BL2018}. Adopting standard models for the gaseous disk and its solid content, we quantify the impact of  the dust torque for a wide range of conditions describing the disk/planet system. We show that the total torque can be positive and revert inward planet migration for planetary cores with $M_{\rm p} \lesssim 10 M_\oplus$. We compute formation tracks for low-mass embryos for conditions usually invoked when modeling planet formation processes. Our most important conclusion is that dust torques can have a significant impact on the migration and formation history of planetary embryos. The most important implications of our findings are: $\it{i})$ For nominal dust-to-gas mass ratios $\epsilon \simeq 0.01$, low-mass planets migrate outwards beyond the water ice-line if most of the mass in solids is in particles with Stokes numbers St $\simeq 0.1$. $\it{ii})$. For $\epsilon \gtrsim 0.02-0.05$, solids with small Stokes numbers, St $\simeq 0.01$, can play a dominant role if most of the mass is in those particles. $\it{iii})$ Dust torques have the potential to enable low-mass planetary cores formed in the inner disk to migrate outwards and act as the seed for massive planets at distances of tens of au.
\end{abstract}



\section{Introduction} 
\label{sec:intro}

As planets grow and evolve in the protoplanetary disks where they form, they modify the environment surrounding them, which in turn can alter the planet's evolution.  In particular, a planet embedded in a disk leads to asymmetric structures with the ability to exert a net torque on the planet that can alter its dynamics. The ensuing planet migration is a key process in planet formation \citep[see][and references therein]{Venturini20Review}. 
While the presence of disk solids is widely recognized as critical in many processes involved in planet formation and evolution, its role has not been assessed in planet migration. Building on the first systematic measurements of dust torques on an embedded planet reported by \citet[][hereafter BLP18]{BL2018}, this paper is a first attempt to address this important issue. 

In sufficiently viscous disks, the gas torque exerted onto the planet has two important contributions: the differential Lindblad torque and the corotation torque. The former is negative for typical protoplanetary disk models while the latter depends sensitively on the vortensity/entropy gradients within the planet's corotation region. For low-mass planets, $M_{\rm p} \lesssim 10 M_\oplus$, these two components of the gas torque are of the same order of magnitude but have opposite signs, leading to the so-called type-I migration regime \cite[e.g.,][]{Tanaka02}. Positive corotation torques can dominate over the differential Lindblad torque locally -- so that a planet migrates outwards transiently \citep[e.g.,][]{Dittkrist2014, Baillie2016, Guilera2017b}. However, overall, planets tend to experience fast inward migration on a global scale. This makes it challenging to explain the existence of giant planets at moderate-to-large distances from the central star \citep[e.g.,][]{Miguel2011, Ronco2017, Voelkel2022}. 
If planetary cores of the order of $10 M_{\oplus}$ can hover around these regions while there is still sufficient gas in the disk, this may lend a possible way to form giant planets at these locations. 

In order to alleviate the fast-inward migration problem, different mechanisms have been proposed to slow down -- or even revert -- the migration of growing planets. Most of these focus on thermodynamic effects in non-isothermal disks or modifications of temperature/entropy and gas density gradients close to the planet \citep[see][for a recent review, and references therein]{Paardekooper2022}. In addition, \citet{Guilera2017} and \citet{Guilera20} show that pressure maxima induced by a viscosity transition in the disk can act as a migration trap for growing planets. Moreover, \citet{Benitez-llambay2015} show that the luminosity of an accreting planet is able to modify the temperature/density close to the planet asymmetrically in a way that an additional ``heating torque'' develops. The heating torque is always positive and can slow down or even revert the planet's migration. More recently, \citet{Guilera2019} and \citet{Guilera2021} incorporated this effect into a global model of planet formation, showing that heating torques can generate significant outward migration for planets growing by pebble accretion. 

Given the fact that protoplanetary disks are made of gas and a small fraction of solids --elements condensed as grains or dust-- it is relevant to consider the possibility that gravitational torques exerted by solids can affect planet migration. Perhaps because solids are thought to contribute with about only 1\% of the total mass of the protoplanetary disk, there have only been a handful of studies about the exchange of angular momentum between the planets and the solid component of the disk \citep[e.g.,][]{Capobianco2011, Chrenko2017, BL2018, Chen2018, Kanagawa2019, Pierens2019, Regaly2020}. In particular, the 2D hydrodynamical simulations from BLP18 show that for low- to intermediate-mass planets, an asymmetric dust-density distribution develops and exerts a net torque onto the planet. This ``dust torque'' is in general positive and its value depends on the Stokes number of solids, the dust-to-gas mass ratio, and the planet mass. It can be of the order (or much larger) than the gas torque and can locally slow down, halt, or revert planet migration for planetary cores with $M_{\rm p} \lesssim 10 M_{\oplus}$. It is then imperative to consider the cumulative effect of the dust torque when integrating the equations of motion of the embedded forming planets.

The aim of this work is to build on the torque measurements provided by BLP18 and investigate the global dynamics of low-mass planetary cores embedded in dusty disks in order to quantify the impact of the dust torque on their migration and formation history. The rest of the paper is organized as follows. In \S\ref{sec:torque_definitions}, we present the scope of our framework for computing the torque exerted on a planet embedded in a gaseous disk that can contain a dust component. Subsequently, we compute torque maps, characterizing the dependence of the total torque on a planet on the various physical parameters describing the planet-disk system and/or the dust component. These include the planet's location, the dust-to-gas mass ratio, the stellar mass accretion rate, and the dust Stokes number distribution. We present the results for steady-state locally-isothermal disks (with a radial temperature gradient) and non-isothermal disks (with radial and vertical thermal structure) in \S\ref{sec:FiducialModel} and \S\ref{sec:non-isothermal disk}, respectively. Based on these torque maps, in \S\ref{sec:planet_migration}, we compute for the first time the formation tracks of low-mass planets embedded in dusty disks where the torque exerted by the dust component is considered.  This allows us to quantify the role of dust in low-mass planetary migration considering a range of solid-accretion rates on single planets embedded in both steady-state and evolving disks. In \S\ref{sec:discussion} we outline the implications of our findings, together with the most important present caveats and possible directions to address them.  Finally, in \S\ref{sec:takeways} we summarize the most important takeaways and conclusions from this first study.

\section{Framework for Computing Planetary Torques}
\label{sec:torque_definitions}

Computing the torque exerted on an embedded planet by the gas and dust component in a protoplanetary disk requires solving the associated multi-species hydrodynamic equations for a particular disk model, planetary mass, and a dust-size distribution. This is, in all generality, a daunting task. Here, we approach this problem in a simplified way by considering existing models for the gas torque responsible for type-I migration \citep{Tanaka02,jm2017} together with the measurements of the dust-to-gas torque ratio provided by BLP18 obtained for wide range of planetary masses and dust Stokes numbers for a single planet embedded in a standard $\alpha$-disk model in steady-state.

\subsection{Torque Calculation}
\label{sec:TorqueCalculation}

We consider the total torque, $\Gamma_{\text{tot}}$, acting on a planet as given by\footnote{For simplicity, in this first study, we do not include the so-called thermal torque \citep[as in][]{Guilera2021} that could lead to either outward or faster inward migration \citep{masset2017}.}
\begin{equation}
\Gamma_{\text{tot}} = \Gamma_{\text{g}} + \Gamma_{\text{d}}\,,
\end{equation}
where $\Gamma_{\text{g}}$ and $\Gamma_{\text{d}}$ are the torques exerted by the gas and dust, respectively. For a given disk model, these torques are obtained as described below.

\begin{deluxetable*}{c || cccccccccc}[t]
\tabletypesize{\scriptsize}
\tablenum{1}
\tablecaption{Values for $\Gamma_{\rm d}/\Gamma_0$ and $\Gamma_{\rm g}/\Gamma_0$ provided in BLP2018 (see their Fig.~2). \label{table:torques}}
\tablewidth{0pt}
\tablehead {\diagbox{$\rm St$ }{$M_{\rm p}/M_{\oplus}$}  &0.333 &0.486 &0.709 &1.03 &1.51 &2.2 &3.22 &4.69 &6.85 &10.0 }
\decimals
\startdata
0.010 &0.283 &0.309 &0.264 &0.201 &0.142 &0.093 &0.056 &0.029 &0.006 &-0.010\\ 
0.014 &0.371 &0.394 &0.415 &0.333 &0.239 &0.160 &0.098 &0.053 &0.017 &-0.008\\ 
0.021 &0.739 &0.720 &0.618 &0.475 &0.382 &0.276 &0.172 &0.095 &0.041 &-0.002\\ 
0.030 &1.638 &1.332 &1.102 &0.851 &0.625 &0.434 &0.275 &0.185 &0.087 &0.021\\ 
0.043 &3.254 &2.614 &1.961 &1.476 &1.078 &0.757 &0.503 &0.316 &0.174 &0.085\\ 
0.062 &5.776 &4.579 &3.428 &2.461 &1.790 &1.266 &0.850 &0.546 &0.326 &0.167\\ 
0.089 &9.672 &7.297 &5.313 &3.760 &2.620 &1.872 &1.301 &0.880 &0.562 &0.304\\ 
0.127 &13.684 &10.191 &7.337 &5.145 &3.471 &2.374 &1.725 &1.209 &0.805 &0.483\\ 
0.183 &16.058 &11.276 &7.866 &5.231 &3.178 &1.372 &1.740 &1.493 &1.073 &0.701\\ 
0.264 &-1.521 &-18.135 &-8.446 &-0.425 &1.971 &1.746 &2.237 &1.823 &1.351 &0.910\\ 
0.379 &-32.124 &-7.348 &2.307 &4.517 &4.752 &3.232 &2.803 &2.108 &1.499 &1.016\\ 
0.546 &5.523 &8.831 &9.054 &7.882 &5.860 &3.741 &2.716 &1.953 &1.398 &0.986\\ 
0.785 &10.492 &9.803 &7.187 &5.859 &4.314 &3.278 &2.465 &1.821 &1.326 &0.943\\ 
1.129 &11.861 &9.572 &7.850 &5.807 &4.538 &3.367 &2.463 &1.799 &1.299 &0.916\\ 
\hline
\hline
{$\Gamma_{\rm g}/\Gamma_0$} & -3.57 & -3.42 &  -3.25 & -3.05 & -2.82 & -2.55 & -2.25 & -1.95 & -1.68 & -1.50 \\
\enddata
\tablecomments{Values obtained for a reference radius $r_0=1$ au and dust-to-gas mass ratio $\epsilon_0 = 0.01$.}
\end{deluxetable*}

\subsubsection{Gas Torque}
\label{gaseous-torque}

For the locally-isothermal disks discussed in \S\ref{sec:FiducialModel}, we employ the gas torque model associated with type-I migration as derived by \citet{Tanaka02} assuming a globally isothermal disk with a power-law density profile
\begin{equation}
\Gamma_{\text{g}} = -\left[1.364-0.54 \left(\frac{d\ln\Sigma_{\text{g}}}{d\ln r}\right)\right] \Gamma_{\text{p}} \,.
\label{Gamma_I}
\end{equation}
Here, $\Gamma_{\text{g}}$ is the sum of the differential Lindblad and corotation torques and $\Gamma_{\text{p}}$ stands for the reference torque evaluated at the radial location of the planet $r_{\text{p}}$ as given by
\begin{equation}
\Gamma_{\text{p}}= q^2 h_{\rm p}^{-2} \Sigma_{\text{g,p}} r_{\text{p}}^4 \Omega_{\text{K,p}}^2 \,,
\label{eq:Gamma_p}
\end{equation}
with $\Sigma_{\rm g}$ the gas surface density, $q=M_{\text{p}}/M_\star$ the planet-to-star mass ratio, and $h = c_{\rm s}/(r \Omega_{\rm K})$ the disk aspect ratio. Here, $c_{\rm s}$ is the sound speed and $\Omega_{\rm K} = \sqrt{GM_\star/r^3}$ is the Keplerian frequency. The subscript ``p'' indicates quantities being evaluated at $r=r_{\rm p}$. 

Using Eq.~(\ref{Gamma_I}) to compute the gas torque for disks with radial temperature gradients is of course an approximation but this enables us to capture the characteristic dependencies of type-I migration with the underlying disk model in a simple way. 
Note that, under the stated assumptions, $\Gamma_{\text{p}}$ encapsulates all the radial dependencies of the physical parameters characterizing the \emph{gas} torque in a dust-free disk.

For the non-isothermal disk models (i.e., with vertical thermal structure) presented in \S\ref{sec:non-isothermal disk}, we employ the gas torque models responsible for type-I migration derived by \citet{jm2017}. The gas torque $\Gamma_{\rm g}$ we consider is then again the sum of the differential Lindblad and corotation torques and is also proportional to $\Gamma_{\text{p}}$ defined in Eq.~(\ref{eq:Gamma_p}). In this case, the differential Lindblad torque depends mainly on the local gradients of the gas surface density and mid-plane temperature while the corotation torque has four contributions, three of them associated with the disk radial gradients of vortensity, entropy, and temperature, and a fourth one accounting for the viscous production of vortensity, see \cite{jm2017} and Eqs.~(26)--(28) and Eqs.~(A.10)--(A.23) in \cite{Guilera2019}\footnote{We note that what we define as $\Gamma_{\text{p}}$ in Eq.~(\ref{eq:Gamma_p}) in this work is defined as $\Gamma_0$ in \citet[][Eq.~(21)]{Guilera2019} and in BLP2018.}.

\begin{figure}[t!]
    \centering
    \includegraphics[width=1.\columnwidth]{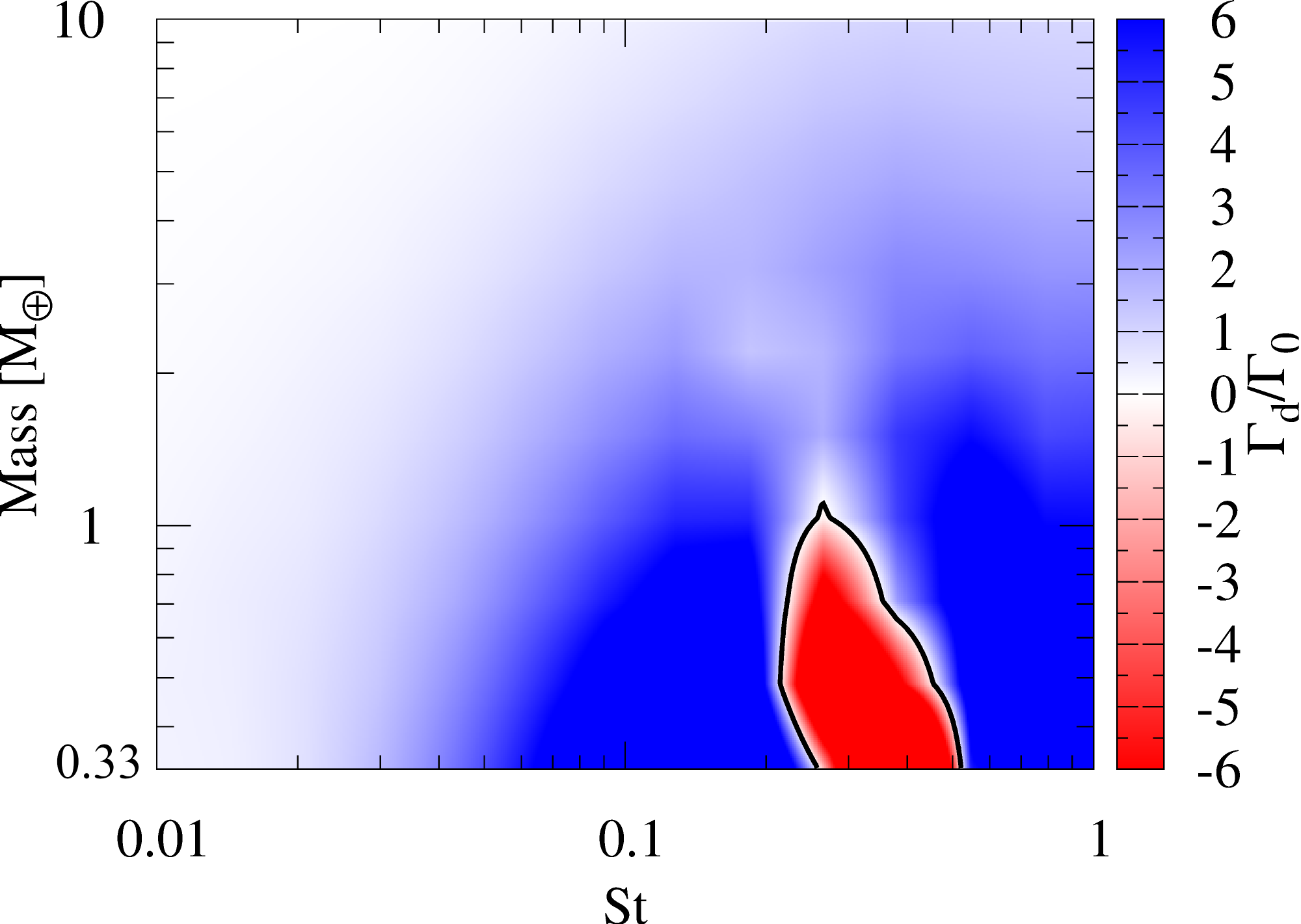} 
    \caption{Normalized reference dust torque, $\Gamma_{\rm d} / \Gamma_{0}$, for a range of planet masses and Stokes numbers for a vertically-isothermal disk as obtained from the simulations in BLP18, with $r_{\rm p} = r_0 = 1$~au and $\epsilon = \epsilon_0 = 0.01$. The values shown correspond to the bilinear interpolation of data in Table \ref{table:torques}.}.
    \label{fig:dust_torque}
\end{figure}

\subsubsection{Dust Torque}
\label{dust-torque}
If the dust-to-gas mass ratio $\epsilon$ is small, we can neglect the dust back-reaction force onto the gas and $\Gamma_{\rm d}$ scales linearly\footnote{For small enough total dust-to-gas mass ratio, the linear scaling holds for each individual dust species.} with $\epsilon$, i.e.,
\begin{equation}
\Gamma_{\rm d}(r_{\rm p}, \epsilon)
 = \frac{\epsilon}{\epsilon_0} \times \Gamma_{\rm d}(r_{\rm p}, \epsilon_0) \,.
 \label{eq:Gamma_e_to_e0}
\end{equation}
We furthermore assume that, similarly to the models for the gas torque\footnote{We note that the ratio $\Gamma_{\rm d}/\Gamma_{\rm p}$ may have an additional dependence with $h$ which we assume to be weak enough to be neglected. Testing the validity of this hypothesis requires dedicated hydrodynamical simulations which are beyond the scope of this paper.}, 
the dust torque (for a given dust-to-gas mass ratio) is proportional to the reference torque, i.e., 
\begin{equation}
\Gamma_{\rm d}(r_{\rm p}, \epsilon_0)
 = \frac{\Gamma_{\rm p}}{\Gamma_{0}} \times \Gamma_{\rm d}(r_0, \epsilon_0) \,.
 \label{eq:Gamma_rp_to_r0}
\end{equation}
where $\Gamma_{\rm p}$ and $\Gamma_0$ stand for the reference torque in Eq.~(\ref{eq:Gamma_p}) evaluated at the location of the planet $r_{\rm p}$ and at the reference location $r=r_0$, respectively.
Combining Eqs.~(\ref{eq:Gamma_e_to_e0}) and (\ref{eq:Gamma_rp_to_r0}), the dust torque can thus be written as
\begin{equation}
\Gamma_{\rm d}(r_{\rm p}, \epsilon)
 = 
 \frac{\epsilon}{\epsilon_0} 
 \times
 \frac{\Gamma_{\rm d}(r_0, \epsilon_0)}{\Gamma_0} \times \Gamma_{\rm p} \,.
\end{equation}
The value of the ratio $\Gamma_{\rm d}(r_0, \epsilon_0)/\Gamma_0$ is obtained from BLP18, where $r_0 = 1$~au and $\epsilon_0 = 0.01$ (see their Fig. 2). The dust torque can then be finally expressed as 
\begin{equation}
\Gamma_{\rm d} = \left(\frac{\epsilon}{0.01}\right) \left( \frac{\Gamma_{\rm d}}{\Gamma_0}\right)_{{\rm BLP18}} \Gamma_{\rm p}\,,
\label{eq:dust_torque}
\end{equation}
where $\left(\Gamma_{\rm d}/\Gamma_0\right)_{{\rm BLP18}}$ is the dust torque measured from numerical simulations in BLP18. These values are listed in Table \ref{table:torques} for logarithmically spaced planetary masses and Stokes numbers, $\text{St}$, in the range $0.1 \le M_{\text{p}}/M_{\oplus} \le 10$ and $10^{-2} \le \text{St} \le 1$, respectively. 

To obtain values of $\Gamma_{\rm d}$ for planetary masses and Stokes numbers that are not tabulated, we use bilinear interpolation in logarithmic space. In Fig.\,\ref{fig:dust_torque}, we show the interpolated dust torques from Table \ref{table:torques}, which have been calculated considering 
$r_{\rm p} = r_0 = 1$~au and $\epsilon = \epsilon_0 = 0.01$.

In order to use Eq.\,\eqref{eq:dust_torque} to model planets migrating and growing in disks with more complex structure, we implicitly assume that $\Gamma_{\rm d}$ does not depend strongly on the gas dynamics or the state of motion of the planet. We note here that the dust torque is expected to be more sensitive to the disk dynamics for small Stokes numbers, for which the dust torque decreases. Also, the dust torque could be sensitive to the state of motion of the planet if the relative drift between solids and the planet is highly affected. This could happen, for example, if migration is very fast (with respect to the radial drift). We also note that the accretion of dust, i.e., the removal of solids inside the planetary Hill sphere, can affect the magnitude of the dust torque \citep[e.g.,][]{Regaly2020}. The validity and robustness of these hypotheses need to be assessed through dedicated hydrodynamical studies. 

\section{Locally Isothermal $\alpha$-disks}
\label{sec:FiducialModel}

We first consider a steady-state, locally isothermal $\alpha$-disk model \citep{SS73} with a dust component characterized by a constant Stokes number throughout the disk as used by BLP18 and described in \citet{Weber2018}. The $\alpha$-parameter relates the accretion rate to the gaseous disk surface density via
\begin{equation}
\Sigma_{\text{g}}= \dfrac{\dot{M}_{\star}}{3 \pi \nu}, 
\label{Sigma_Mdotstar_nu}
\end{equation}
where $\dot{M}_{\star}$ is the mass accretion rate onto the central star, 
$\nu= \alpha c_{\text{s}} h $ is the disk viscosity, and the aspect ratio, $h$, is considered to be constant.  Under these assumptions, the background gas surface density and temperature are power laws in radius, with
\begin{equation}
\left.\frac{d\ln\Sigma_{\text{g}}}{d\ln r}\right|_{\text{isoth.}} = -\frac{1}{2}
\quad
\text{and}
\quad
\left.\frac{d\ln T}{d\ln r}\right|_{\text{isoth.}} = -1 \,.
\end{equation}

\subsection{Torque Maps for Fiducial, Isothermal Disk Model}
\label{sec2.4}

\begin{figure}[t!]
  \centering
  \includegraphics[width=1.\columnwidth]{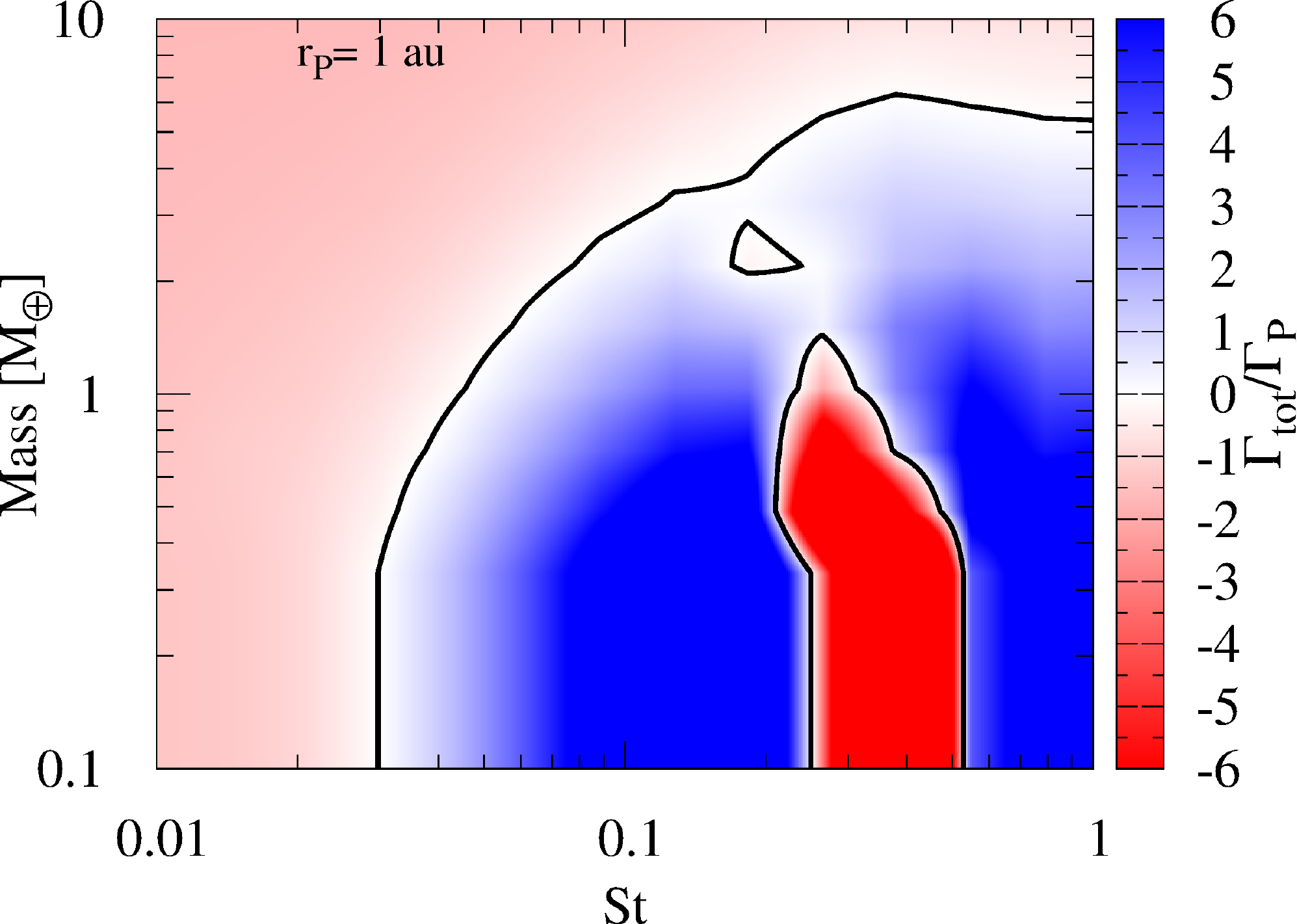} 
  \caption{Normalized torque map, $\Gamma_{\text{tot}}/\Gamma_{\text{p}}$, for a range of planet masses and Stokes numbers at 1 au for a vertically-isothermal disk.  The color scale accentuates the regions where the total torque changes sign. Red/blue tones indicate negative/positive total torque, which locally implies inward/outward planet migration.}
    \label{fig2_sec2.4}
\end{figure}

\begin{figure*}[t!]
    \centering
    \includegraphics[width=1.\textwidth]{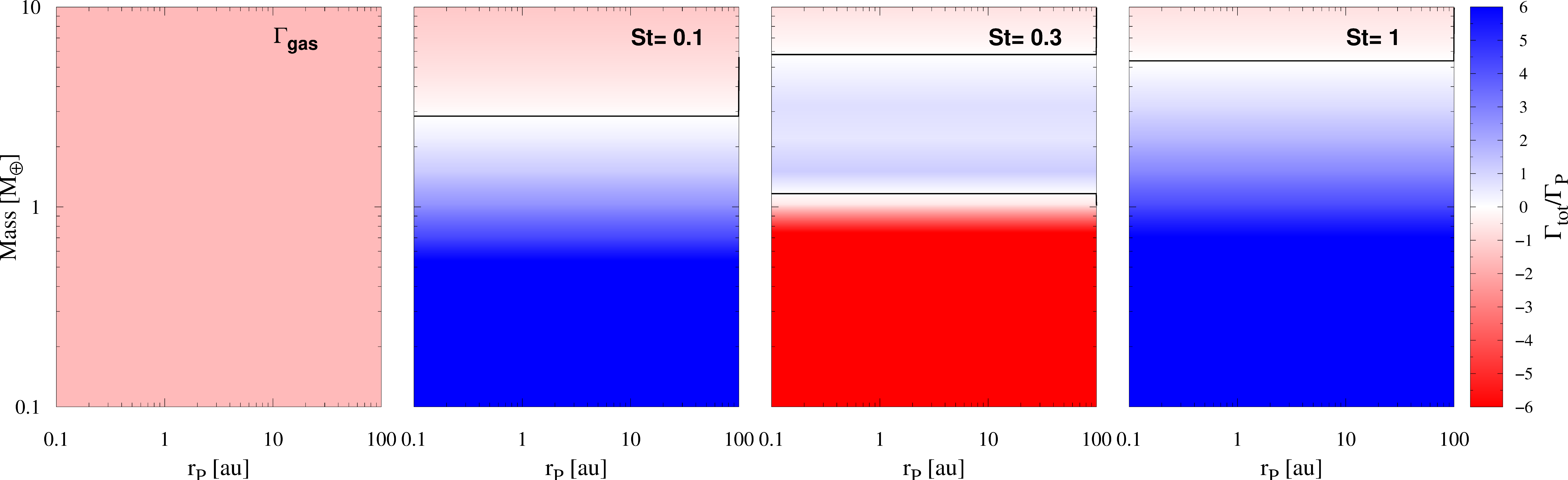} 
    \caption{Normalized torque maps, $\Gamma_{\text{tot}}/\Gamma_{\text{p}}$, for a range of planet masses and locations in an isothermal disk. The left panel corresponds to a dust-free disk, whereas the other panels illustrate the results for a disk with dust-to-gas mass ratio $\epsilon = 0.01$ and particles with constant Stokes number St= \{0.1, 0.3, 1\}. Since for isothermal $\alpha$-disks all the radial dependencies of the torque are encapsulated in $\Gamma_{\text{p}}$, the normalized torques do not depend on the planet's location.}
    \label{fig:torque_map_mp_vs_rp_iso}
\end{figure*}

In what follows, we will employ the term ``torque map'' to characterize the dependence of the total torque $\Gamma_{\text{tot}}$ on the various physical parameters describing the planet-disk system and/or the dust component. It is convenient to present normalized values $\Gamma_{\text{tot}}/\Gamma_{\text{p}}$ using the reference torque $\Gamma_{\rm p}$, defined in Eq.\,\eqref{eq:Gamma_p}.

In order to understand the scales involved in the problem at hand, and build intuition before considering more elaborate models, we first consider dusty disk models similar to those in BLP18. These correspond to steady-state, vertically-isothermal $\alpha$-disks with constant aspect ratio $h= 0.05$ around a central star of $1~\text{M}_{\odot}$ with $\dot{M}_{\star}= 10^{-7}~\text{M}_{\odot}/\text{yr}$ and $\alpha= 3 \times 10^{-3}$. When the dust component is present, we set its constant gas-to-dust mass ratio to $\epsilon = 0.01$.  

The normalized torque map $\Gamma_{\text{tot}}/\Gamma_{\text{p}}$ as a function of planet mass and Stokes number, computed at 1 au for the isothermal disk model considered, is shown in Fig.~\ref{fig2_sec2.4}. The regions of the torque map corresponding to negative/positive values are associated with inward/outward migration. The two distinct regimes identified by BLP18 are evident; the gas-dominated regime for Stokes numbers ${\rm St}\simeq0.05$\textendash$ 0.2$ and planet masses $M_{\text{p}}\lesssim 3~\text{M}_\oplus$, and the gravity-dominated regime for ${\rm St}\gtrsim 0.3$\textendash$ 0.5$ and planet masses $M_{\text{p}}\lesssim 7~\text{M}_\oplus$. 

In the gas-dominated regime (low values of St) the torque increases with the Stokes number, allowing the outward migration of more massive planets as St increases (up to a masses of $3~\text{M}_\oplus$ at $\text{St}\simeq 0.1$). On the other hand, in the gravity-dominated regime, the total torque is less sensitive to the value of St, and the maximum mass for outward migration stays at about $M_p\simeq 5~\text{M}_\oplus$ to  $6~\text{M}_\oplus$. In between these two regimes there exists a transition region (${\rm St}\simeq 0.2$\textendash$ 0.5$) where the intensity of the dust torque is strongly reduced and can even change sign resulting mostly in inward migration. The total torque in this transition regime can depend sensitively on the mass of the planet. For the standard value of the dust-to-gas mass ratio considered here, $\epsilon=0.01$, the total torque is practically equal to the gas torque for low Stokes numbers (${\rm St}\lesssim0.02$). However, as we will show in \S~\ref{sec:dust_to_gas_ratio} this situation can drastically change if the dust-to-gas mass ratio increases.

The normalized torque map $\Gamma_{\text{tot}}/\Gamma_{\text{p}}$  as a function of planet mass and disk location associated with our fiducial disk model is shown in Fig.~\ref{fig:torque_map_mp_vs_rp_iso}. The leftmost panel corresponds to a dust-free disk whereas the subsequent panels show the results for dusty disks with three different (constant) Stokes numbers St= \{0.1, 0.3, 1\}.
For the dust-free disk, $\Gamma_{\text{tot}} = \Gamma_{\text{g}}$ and $\Gamma_{\text{tot}}/\Gamma_{\text{p}}$ is, by construction, constant with 
$\Gamma_{\text{tot}}/\Gamma_{\text{p}}= -1.634$. In the cases corresponding to St= 0.1 and St= 1, the dust torque generates a significant region of positive total torque, depicted with blue. For St= 0.1, this positive total torque region extends up to $\simeq 3~\text{M}_{\oplus}$, while for the case of St= 1 the positive total torque region extends up to $\simeq 5~\text{M}_{\oplus}$. The situation is quite different for the case of St= 0.3 where outward migration only ensues for masses $1~\text{M}_\oplus \lesssim M_{\text{p}} \lesssim 6~\text{M}_\oplus$ and becomes inward outside that range. 

Note that for the isothermal disk model under consideration, all radial dependencies are encapsulated in 
$\Gamma_{\text{p}}$ and the value of the total torque depends only on the mass on the planet and not on its location. Furthermore, because the disk aspect ratio is known, in this case $h= 0.05$, $\Gamma_{\text{p}}$ can be directly computed given the planet mass and planet location, and the total torque can be easily reconstructed from the color magnitude of the maps. In the next section, we will see that for non-isothermal disks (where the disk aspect ratio $h$ is no longer constant), the transition between positive and negative total torques depends on the planet location; with implications for its migration history.

Having gained some insight into the relative contribution of the dust torque in the framework of the isothermal $\alpha$-disk, we consider next a more realistic disk model.

\section{Non-isothermal $\alpha$-disks}
\label{sec:non-isothermal disk}

We consider a steady-state, non-isothermal axisymmetric $\alpha$-disk with radial and vertical thermal structure. The disk model is obtained by solving the classical structure and transport equations in the vertical direction, considering that vertical hydrostatic equilibrium at each radius results from viscous heating and irradiation from the central star\footnote{We used $\text{T}_{\text{eff}}= 4397$~K and $\text{R}_{\star}= 3.079~\text{R}_{\odot}$ for a proto-star of $1~\text{M}_{\odot}$ as provided in \citet{Baraffe15}.}. Specifically, in order to compute the vertical structure of the disk, for a given $\alpha$-viscosity parameter and stellar mass accretion rate $\dot{M}_{\star}$, we solve at each radial location the set of equations
\begin{subequations}
\begin{equation}
\frac{\partial P}{\partial z} = -\rho \Omega^2 z,
\end{equation}
\begin{equation}
\frac{\partial F}{\partial z} = \frac{9}{4} \rho \nu \Omega^2, 
\label{eq1-sec2-1}
\end{equation}
\begin{equation}
\frac{\partial T}{\partial z} = \nabla\frac{T}{P}\frac{\partial P}{\partial z}, 
\end{equation}
\end{subequations}
where $P$, $\rho$, $\Omega$, $T$, $F$ and $z$ are the pressure, density, angular frequency, temperature, heat flux, and vertical distance from the disk midplane, respectively, and $\nabla= \text{d}\ln\,T/\text{d}\ln\,P$ \citep[see][for further details]{Guilera2017b, Guilera2019}.

\begin{figure}[t!]
  \centering
  \includegraphics[angle= 270, width=1.\columnwidth]{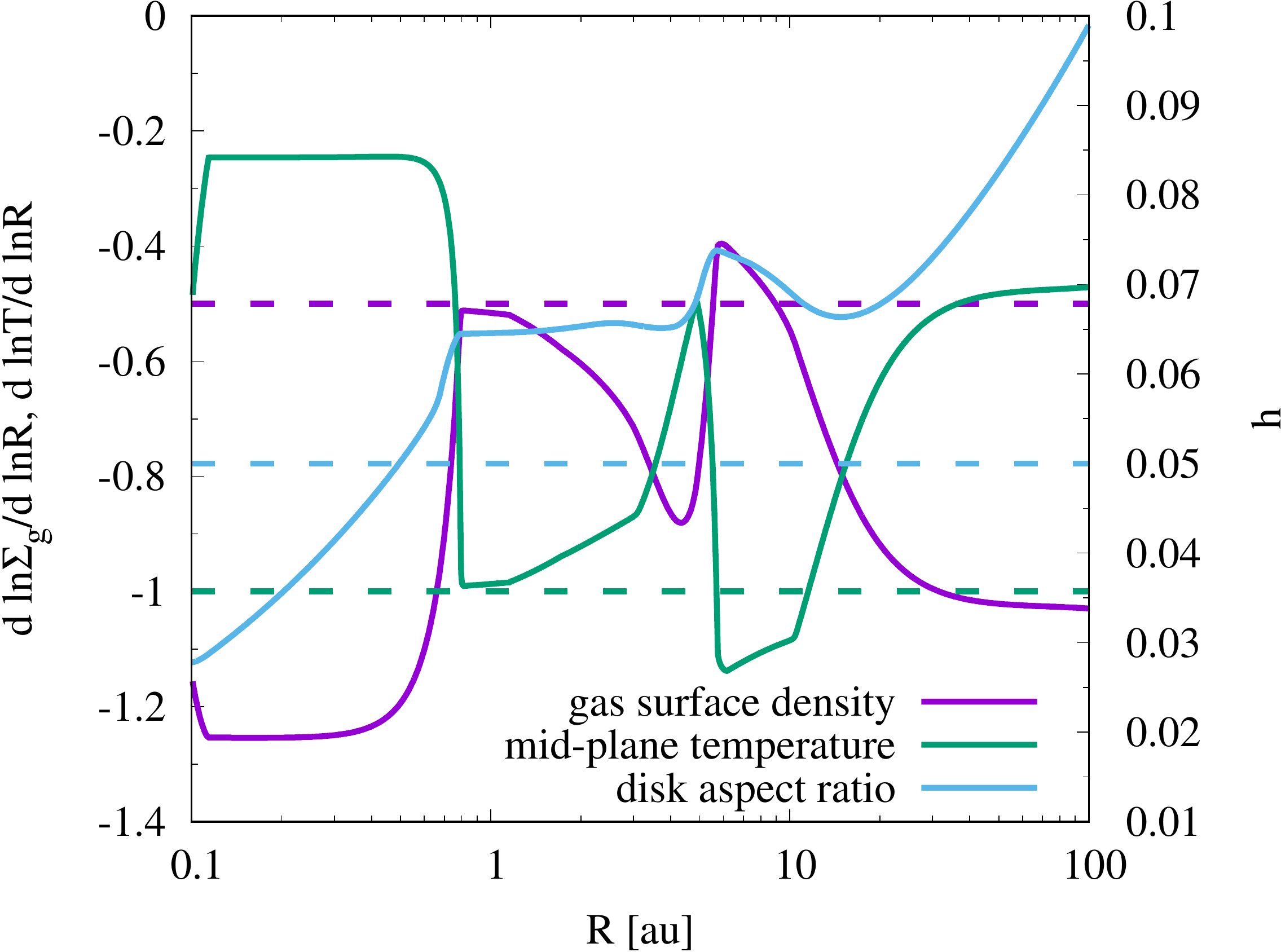} 
  \caption{Radial profiles of the logarithmic gas surface density and mid-plane temperature gradients, and the disk aspect ratio, for the steady-state solution of an $\alpha$-disk considering $\dot{M}_{\star}= 10^{-7}~\text{M}_{\odot}/\text{yr}$ and $\alpha= 3 \times 10^{-3}$. The solid lines correspond to the non-isothermal disk, while the dashed ones illustrate the isothermal disk case. The abrupt changes in the mid-plane temperature and gas surface density gradients are due to sharp disk opacity transitions at such locations.} 
    \label{fig1_sec3}
\end{figure}

The logarithmic gradients for the gas surface density and mid-plane temperature, as well as the disk aspect ratio $h$, for a non-isothermal disk with $\alpha= 3 \times 10^{-3}$ and $\dot{M}_{\star}= 10^{-7}~\text{M}_{\odot}/\text{yr}$ are shown in Fig.~\ref{fig1_sec3}. The (constant) values corresponding to the isothermal disk model are shown with dashed lines in the same figure for comparison. It is evident that the radial structure of the non-isothermal disk model cannot be described by a unique power law. In these models, heat transport and, in particular, radiative opacities play a key role in determining the density and mid-plane temperature profiles of the disk. The non-trivial radial disk structure may lead to noticeable differences between the non-isothermal and isothermal gas torque maps. However, in order to obtain the dust torque for the non-isothermal disk we assume that the dust torque can still be computed using Eq.~(\ref{eq:dust_torque}), where the non-isothermal disk aspect ratio is used when computing the reference torque $\Gamma_{\text{p}}$.

\begin{figure}[t!]
  \centering
  \includegraphics[width=1.\columnwidth]{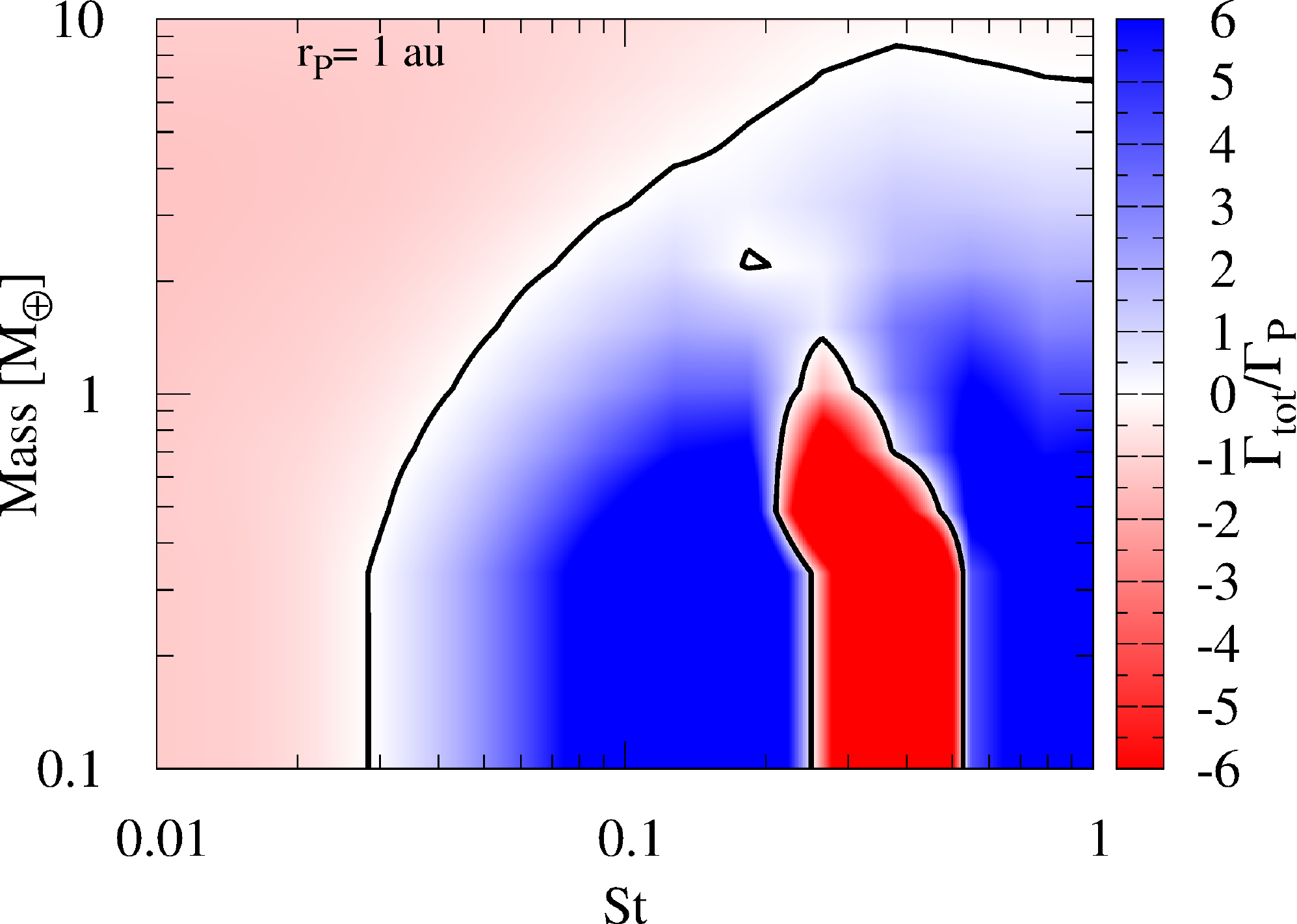} 
  \caption{Same as Fig.~\ref{fig2_sec2.4}, but for the case of the non-ishotermal disk.}
  \label{fig3_sec3}
\end{figure}

The normalized torque map $\Gamma_{\text{tot}}/\Gamma_{\text{p}}$ as a function of planet mass and Stokes number, computed at 1~au for the non-isothermal disk with a dust-to-gas mass ratio $\epsilon=0.01$, is shown in Fig.~\ref{fig3_sec3}. As in the case of the isothermal disk model, the two regions of positive total torque, as well as the transition between these at ${\rm St}\simeq 0.2$\textendash$ 0.5$, are clearly visible. Similarly to the vertically isothermal disk, the dust torque does not play a relevant role for the standard dust-to-gas mass ratio of $\epsilon_0=0.01$ when dust particles with low Stokes numbers e.g., $\text{St} \lesssim 0.02$ are considered. All of these features in the torque maps evaluated at 1~au, as seen in Fig.~\ref{fig3_sec3}, are also broadly present in these torque maps when calculated at other disk locations. However, due to the radial disk structure seen in Fig.~\ref{fig1_sec3}, the value of $\Gamma_{\text{tot}}/\Gamma_{\text{p}}$ does depend on the planet's location, in contrast to the isothermal disk case.

\subsection{Dependence on Planet Location}
\label{}

\begin{figure*}[t!]
    \centering
    \includegraphics[width=1.\textwidth]{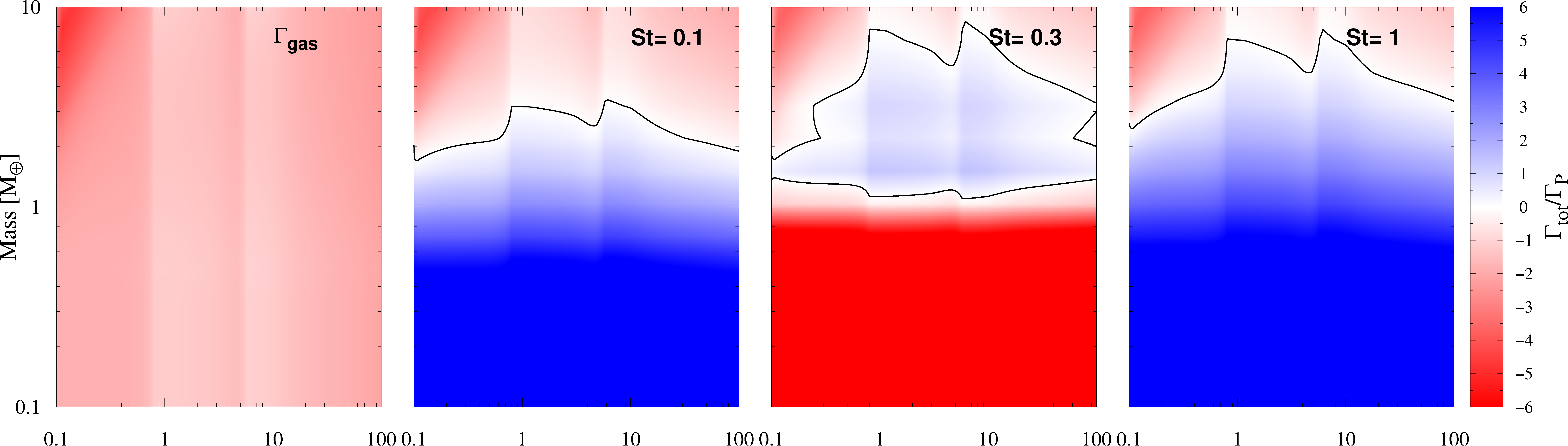} \\
    \centering
    \includegraphics[width=1.\textwidth]{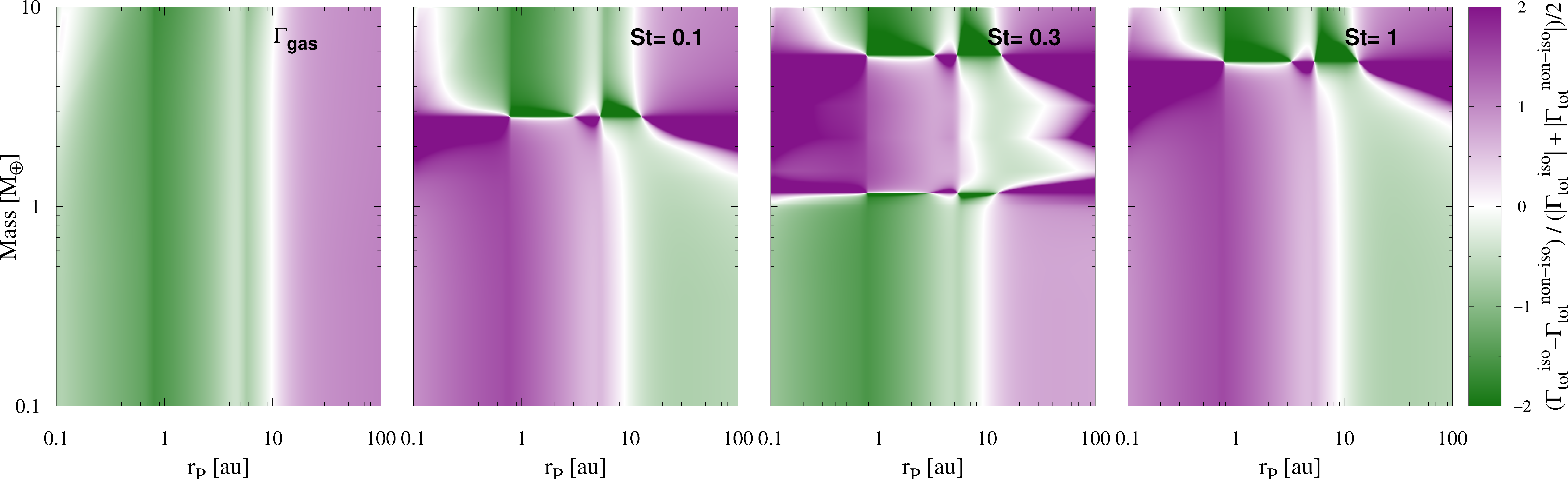}
    \caption{Top row: normalized torque maps, $\Gamma_{\text{tot}}/\Gamma_{\text{p}}$, as a function of planet mass and location for the non-isothermal disk model with $\alpha= 3 \times 10^{-3}$, $\dot{M}_{\star}= 10^{-7}~\text{M}_{\odot}/\text{yr}$, and dust-to-gas mass ratio $\epsilon=0.01$. The disk model is described in detail in \S\ref{sec:non-isothermal disk}. Bottom row: normalized differential torque maps between the isothermal and non-isothermal cases.}
    \label{fig:torque_map_mp_vs_rp_non_iso}
\end{figure*}

The normalized torque map $\Gamma_{\text{tot}}/\Gamma_{\text{p}}$  as a function of planet mass and location for the non-isothermal disk model with gas torque as provided in \citet{jm2017} is shown in the top row of Fig.~\ref{fig:torque_map_mp_vs_rp_non_iso}. The leftmost panel corresponds to a dust-free disk whereas the subsequent panels show the results for dusty disks with three different (constant) Stokes numbers St= \{0.1, 0.3, 1\}. Similarly to the vertically isothermal disk, the total gas torque is always negative, implying inward planet migration. However, in this case the gas torque presents a slightly more complex dependence on the location of the planet (see below).

The results for the dusty disks displayed in Fig.~\ref{fig:torque_map_mp_vs_rp_non_iso} are qualitatively similar to the case of the isothermal disk: for the cases of St= 0.1 and St= 1, the dust torque generates a significant region of positive total torque. In these two cases, differently, than for the isothermal disk case, the value of the planet mass at which there is a transition between negative and positive torque regions depends on the planet's location in the disk, see  Fig.~\ref{fig1_sec3}. This non-trivial radial dependence observed also to a lesser extent in the dust-free disk is due to the fact that the radial structure for the non-isothermal disk deviates from power laws and that the Lindblad and corotation torques as provided in \citet{jm2017} depend on the disk diffusivity and viscosity.

For disks with dust particles with St= 0.3, the result is also qualitatively similar to the vertically isothermal disk. We find a total negative torque throughout the disk for low-mass (up to about $\simeq 1~\text{M}_\oplus$) planets. As the mass of the planet increases (at a fixed location), the total torque becomes positive, but it becomes negative again beyond a certain mass. As explained above, for non-isothermal disks the mass threshold for the transition between regions of positive and negative total torque depends on the location of the planet due to the radial disk structure seen in Fig.~\ref{fig1_sec3}.

In order to illustrate the relative difference between the torque maps obtained for isothermal and non-isothermal disk models considered, the bottom row of Fig.~\ref{fig:torque_map_mp_vs_rp_non_iso} shows the associated differential torque map. The aim of this figure is to show how the radial dependence between both models differs, especially when the dust torque is included in the computation of the total torque. The radial structure of the normalized differential torque map in the dust-free case simply reflects the structure of the normalized non-isothermal gas torque map (the leftmost panel on the top row in Fig.~\ref{fig:torque_map_mp_vs_rp_non_iso}) because the normalized isothermal gas torque map is constant. When dust is present, the radial structure of the differential torque map is more complex. For St= 0.1 and St=1 there is a trend, for low-mass planets wherein the total torque is positive, the total isothermal torque is larger (lower) than the total non-isothermal torque within (beyond) $\sim 10$~au. For larger planets wherein the total torque is negative the trend is the opposite, the total isothermal torque is lower (larger) than the total non-isothermal torque within (beyond) $\sim 10$~au. For the case of St= 0.3 and the regions where the total torque is negative, the total isothermal torque is more (less) negative than the non-isothermal within (beyond) $\sim 10$~au. On the other hand, in the region where the total torque is positive, the total isothermal torque is lower (larger) than the non-isothermal within (beyond) $\sim 10$~au. 

\begin{figure*}[t!]
    \centering
    \includegraphics[width=1.0\textwidth]{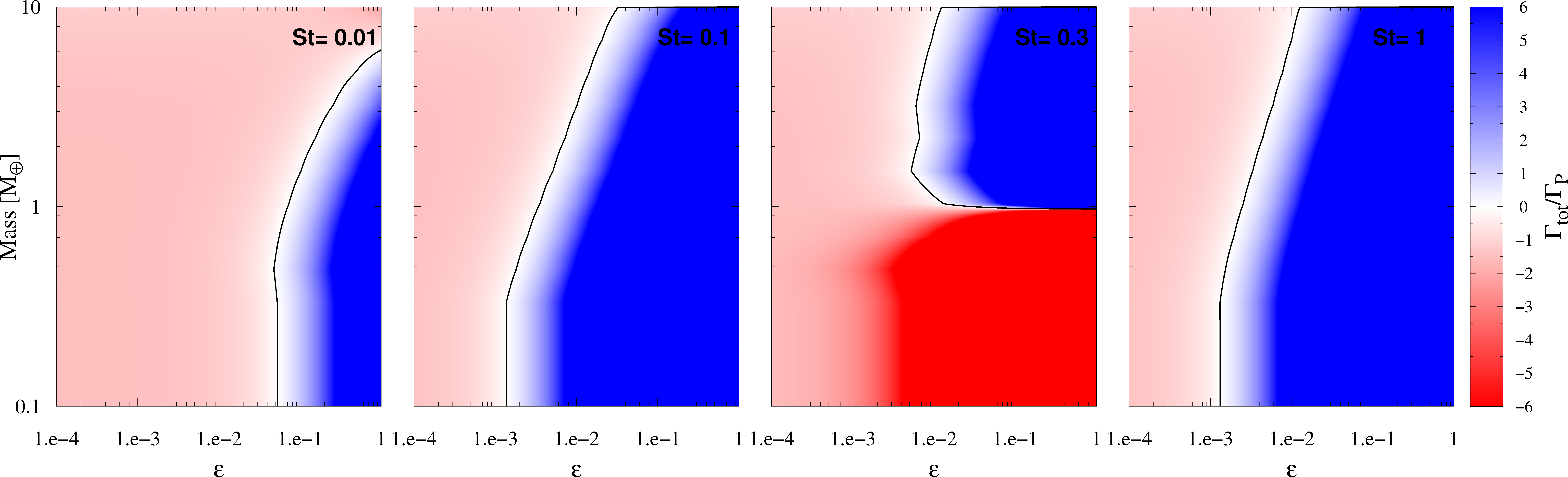} 
    \caption{Normalized torque maps, $\Gamma_{\text{tot}}/\Gamma_{\text{p}}$, as a function of planet mass and dust-to-gas mass ratios computed at 1~au for a non-isothermal disk model with $\alpha= 3 \times 10^{-3}$, $\dot{M}_{\star}= 10^{-7}~\text{M}_{\odot}/\text{yr}$, and dust-to-gas mass ratio $\epsilon=0.01$. The disk model is described in detail in \S\ref{sec:non-isothermal disk}. The panels correspond to different constant Stokes number St= \{0.01, 0.1, 0.3, 1\}, from left to right.}
    \label{fig1_sec4.1}
\end{figure*}

\begin{figure*}[t!]
    \centering
   \includegraphics[width=1.0\textwidth]{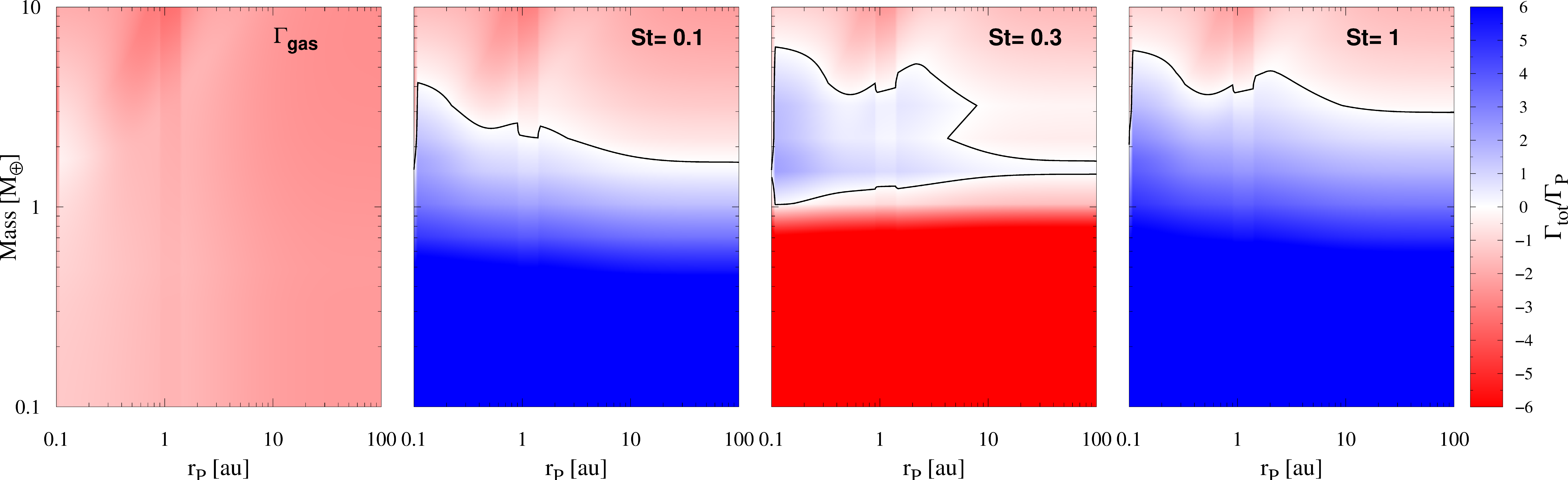} 
    \caption{Same as Fig.~\ref{fig:torque_map_mp_vs_rp_non_iso} but for a steady-state solution of a non-isothermal $\alpha$-disk using $\dot{M}_{\star}= 10^{-9}~\text{M}_{\odot}/\text{yr}$.} 
    \label{fig1_sec3.2}
\end{figure*}

\subsection{Dependence on Dust-to-gas Mass Ratio}
\label{sec:dust_to_gas_ratio}

The value of the dust-to-gas mass ratio can in principle depart significantly with respect to the fiducial value  $\epsilon=0.01$ and it is thus necessary to understand how the total torque on a planet changes accordingly. The normalized torque maps $\Gamma_{\text{tot}}/\Gamma_{\text{p}}$ for the non-isothermal disk as a function of planet mass and dust-to-gas mass ratio, $\epsilon$, for four different Stokes numbers St= $\{0.01, 0.1, 0.3, 1\}$, are shown in Fig.~\ref{fig1_sec4.1}. These torque maps correspond to a planet located at 1~au but the results obtained at other locations in the disk are qualitatively similar.

As seen in Fig.~\ref{fig1_sec4.1}, planets with $M_{\text{p}}\lesssim 0.5~\text{M}_\oplus$ embedded in a disk with dust of small Stokes number St $\simeq 0.01$ can be subject to strong positive torques for dust-to-gas mass ratios $\epsilon \gtrsim 0.05$.  As the dust-to-gas mass ratio increases, the total torque becomes positive for planets with larger masses. Planets with masses in the range $0.5~\text{M}_\oplus$ to $5~\text{M}_\oplus$ can still experience a positive net torque provided the dust-to-gas mass ratio increases from $\epsilon \simeq 0.05$ to $\epsilon \simeq 1$. This dependence of the total torque on the dust-to-gas mass ratio may play an important role in the formation of planets by pebble accretion inside the water ice-line, where Stokes numbers tend to be small (\citetalias[][in preparation]{Guilera22b}).

The morphology of the torque maps for St= 0.1 and St= 1 is qualitatively similar to the case with St = 0.01. However, particles with these Stokes numbers, respectively associated with the gas- and gravity-dominated regimes discussed by BLP18, can exert large positive torques at significantly lower dust-to-gas mass ratios. For instance, planets with $M_{\text{p}}\lesssim 0.3~\text{M}_\oplus$ experience positive total torques for dust-to-gas mass ratios as low as $\epsilon \gtrsim 2 \times 10^{-3}$. Planets of higher masses require higher dust densities to experience positive total torques but it is remarkable that planets as massive as $10~\text{M}_\oplus$ experience positive torques for dust-to-gas mass ratios as low as $\epsilon \gtrsim 2 \times 10^{-2}$. This behavior may have important implications for the formation of planets by pebble accretion outside the water ice-line (see \citetalias[][in preparation]{Guilera22b}).
 
For Stokes numbers in the transition region (St $\simeq 0.3$) the torque map is very different, with positive total torques arising only when $\epsilon \gtrsim 0.01$ and $M_{\text{p}}\gtrsim 1~\text{M}_\oplus$ (at 1~au). Due to the strong dependence of the torques with the location of the planet in this regime (see Fig. \ref{fig:torque_map_mp_vs_rp_non_iso}) we expect the range of masses and dust-to-gas mass ratios for outward migration to be sensitive to the planet's distance to the central star. 

\begin{figure*}[t!]
    \centering
    \includegraphics[width=1.0\textwidth]{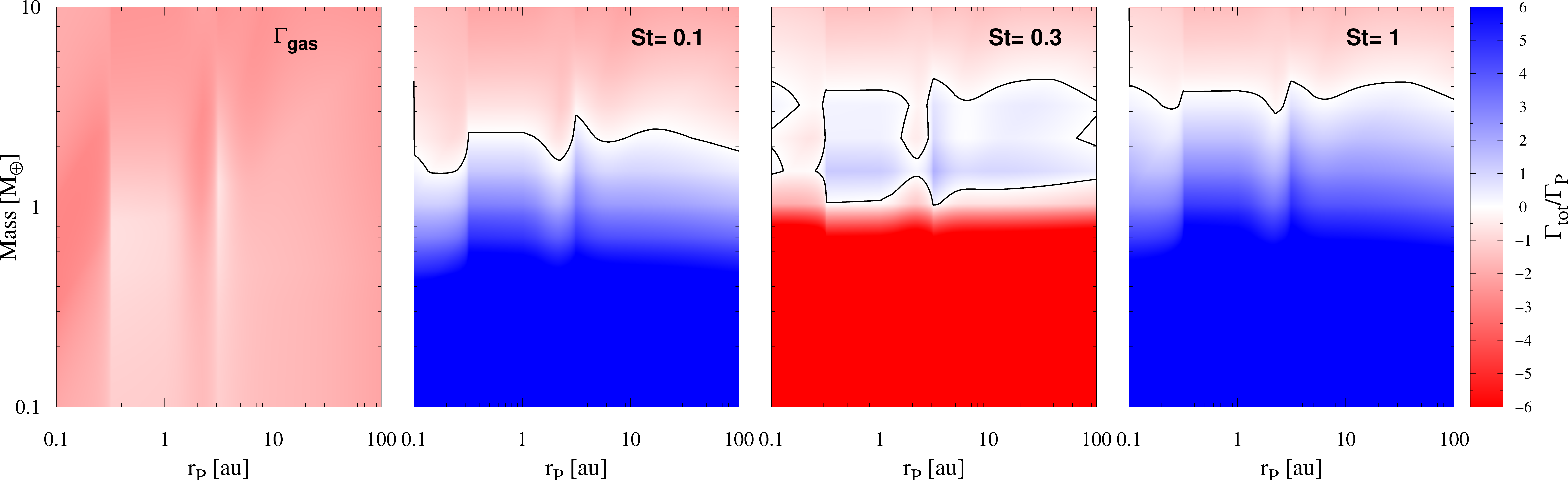} 
    \caption{Same as Fig.~\ref{fig:torque_map_mp_vs_rp_non_iso} but for a steady-state solution of the non-isothermal $\alpha$-disk using $\alpha= 10^{-4}$.} 
    \label{fig1_sec3.3}
\end{figure*}

\begin{figure*}[t!]
    \centering
    \includegraphics[width=1.0\textwidth]{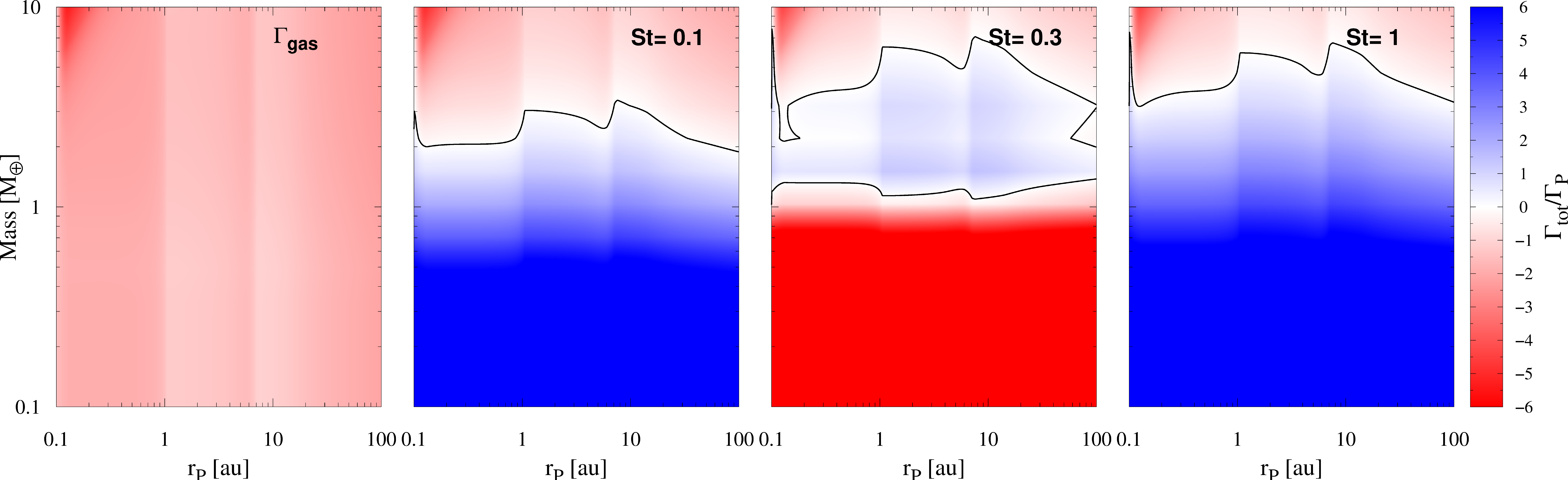}
    \caption{Same as Fig.~\ref{fig:torque_map_mp_vs_rp_non_iso} but for a steady-state solution of the non-isothermal $\alpha$-disk using $\alpha= 10^{-2}$.} 
    \label{fig2_sec3.3}
\end{figure*}

\subsection{Dependence on Mass Accretion Rate} 
\label{sec:dependence_mass_accretion_rate} 

Protoplanetary disks typically evolve on time-scales of the order of $1$ to $10$~Myr. During this evolution, the stellar mass accretion rate $\dot{M}_{\star}$ drops from about $10^{-7}~\text{M}_{\odot}/\text{yr}$ for the youngest proto-stars to $\simeq 10^{-9}~\text{M}_{\odot}/\text{yr}$ for the oldest ones. This reduction in the stellar mass accretion rate is associated with a lower disk gas surface density \citep{Hartmann98}. This motivates the analysis of the dependence of the torque maps on the mass accretion rate $\dot{M}_{\star}$. 

The normalized torque maps as a function of planet mass and location for a non-isothermal disk with $\dot{M}_{\star}= 10^{-9}~\text{M}_{\odot}/\text{yr}$ for a dust-free disk and for a dusty disk with three different (constant) Stokes numbers St= $\{0.1, 0.3, 1\}$ are shown in Fig.~\ref{fig1_sec3.2}.
Note that even though the normalized torques are similar in magnitude to the case with $\dot{M}_{\star}= 10^{-7}~\text{M}_{\odot}/\text{yr}$, the gas torque is about one order of magnitude lower for the lower accretion rate. This reflects the fact that the surface density of $\alpha$-disk models in steady-state decreases with decreasing mass accretion rate $\dot{M}_{\star}$ according to Eq.~(\ref{Sigma_Mdotstar_nu}).

The normalized torque map for the dust-free disk is very similar to the case with $\dot{M}_{\star}= 10^{-7}~\text{M}_{\odot}/\text{yr}$ and the total torque is always negative. For disks with dust particles with Stokes number St= 0.1 the dust torque remains dominant, leading to a positive net torque, for planets with masses up to roughly $\simeq 1.5~\text{M}_{\oplus}$ and $\simeq 4~\text{M}_{\oplus}$. 
For Stokes number St= 1 the total torque remains positive for planets with higher masses, of the order of $\simeq 3~\text{M}_{\oplus}$ to $\simeq 6~\text{M}_{\oplus}$ depending on the planet location. We also note that in these last cases, despite the fact that they present similar trends than those observed for $\dot{M}_{\star}= 10^{-7}~\text{M}_{\odot}/\text{yr}$, the radial dependence of the transition between positive and negative torques is different.  In contrast to the behavior seen in Fig.~\ref{fig:torque_map_mp_vs_rp_non_iso}, for the higher accretion rate $\dot{M}_{\star}= 10^{-7}~\text{M}_{\odot}/\text{yr}$, here the dust torque dominates closer to the central star. 
As the disk evolves and its accretion rate and surface density decrease, dust particles with either St= 0.1 or St= 1 may be relevant for enabling low-mass planets in the inner disk regions to migrate outwards during the late stages of the disk evolution (see $\S$\ref{sec:discussion}).

For St= 0.3, similarly to the case with the accretion rate $\dot{M}_{\star}= 10^{-7}~\text{M}_{\odot}/\text{yr}$,  at a given disk location, there is a range of masses for which the total torque is negative. This mass range decreases from 
$1M_\oplus \lesssim M_{\text{p}}\lesssim 6M_\oplus$ in the inner disk to 
$2M_\oplus \lesssim M_{\text{p}}\lesssim 3M_\oplus$ in the outer disk. At lower accretion rates the total torque is negative for almost all planet masses beyond $r_{\rm p}\gtrsim 7$~au.

\subsection{Dependence on the $\alpha$-viscosity Parameter}
\label{sec:dependence_viscosity}

Here we analyze the dependence of the torque maps on the $\alpha$-viscosity parameter. We repeat the computations in $\S$\ref{sec:non-isothermal disk} considering the values $\alpha= 10^{-2}$ and $\alpha= 10^{-4}$. We note that in order to ease the comparison, we adopt different accretion rates than in the case of $\alpha= 3 \times 10^{-3}$. If we were to use the same value $\dot{M}_{\star}= 10^{-7}~\text{M}_{\odot}/\text{yr}$ the disk gas surface density corresponding to $\alpha= 10^{-4}$ would be much higher than for $\alpha= 3 \times 10^{-3}$.  This is due to the fact that when the viscosity is decreased the gas surface density needed to maintain a constant accretion rate throughout the disk must increase accordingly. The opposite occurs for $\alpha= 10^{-2}$. Thus, in order to achieve similar gas surface density profiles, we adopt $\dot{M}_{\star}= 3 \times 10^{-7}~\text{M}_{\odot}/\text{yr}$ for $\alpha= 10^{-2}$ and $\dot{M}_{\star}= 3 \times 10^{-9}~\text{M}_{\odot}/\text{yr}$ for $\alpha= 10^{-4}$. We note that even though the gas surface density radial profiles are similar in these three cases, the corresponding mid-plane temperatures are higher for larger $\alpha$ values and the local mid-plane temperature gradients differ.   

The normalized torque maps for $\alpha= 10^{-4}$ and $\alpha= 10^{-2}$ are shown in Fig.~\ref{fig1_sec3.3} and Fig.~\ref{fig2_sec3.3}, respectively.
The trends observed are similar to the case with $\alpha= 3 \times 10^{-3}$. For the dust-free disk the total torque is always negative. For St= 0.1 and St=1, the dust toque becomes the dominant contribution of the total torque, generating significant regions of positive torque. In the transition regime, for St= 0.3, we obtain negative total torques for the less massive and more massive planets and positive total torques for intermediate masses. The transition between regimes presents a complex radial dependence, and in the case of $\alpha= 10^{-4}$ the total torque is negative for all masses for planets located at $r_{\rm p}\simeq 0.2$~au. 

It is worth noticing that the main features of the torque maps, including the magnitude of the normalized torques, are rather insensitive to the precise value of $\alpha$, a highly unknown parameter in real protoplanetary disks. 
For $\alpha=10^{-2}$, dust diffusion may have a significant effect, particularly for particles with small Stokes number. However, the contribution of such particles to the dust torque is less significant. Additional hydrodynamical calculations, including the effect of dust diffusion in highly turbulent disks, are required to characterize the dependence of dust torques on viscosity.
 
\subsection{Impact of Dust-size Distribution} 
\label{sec:stokes_number_distri}

In all previous sections, we have considered that the dust component can be characterized by a single Stokes number. Here, we consider dust with a distribution of Stokes numbers. For illustrative purposes, and to ease the comparison with the results in the previous sections, we adopt again a steady-state, non-isothermal $\alpha$-disk with $\alpha= 3 \times 10^{-3}$ and  $\dot{M}_{\star}= 1 \times 10^{-7}~\text{M}_{\odot}/\text{yr}$ and a dust-to-gas mass ratio $\epsilon= 0.01$.

\begin{figure}[t!]
  \centering
  \includegraphics[width=1.\columnwidth]{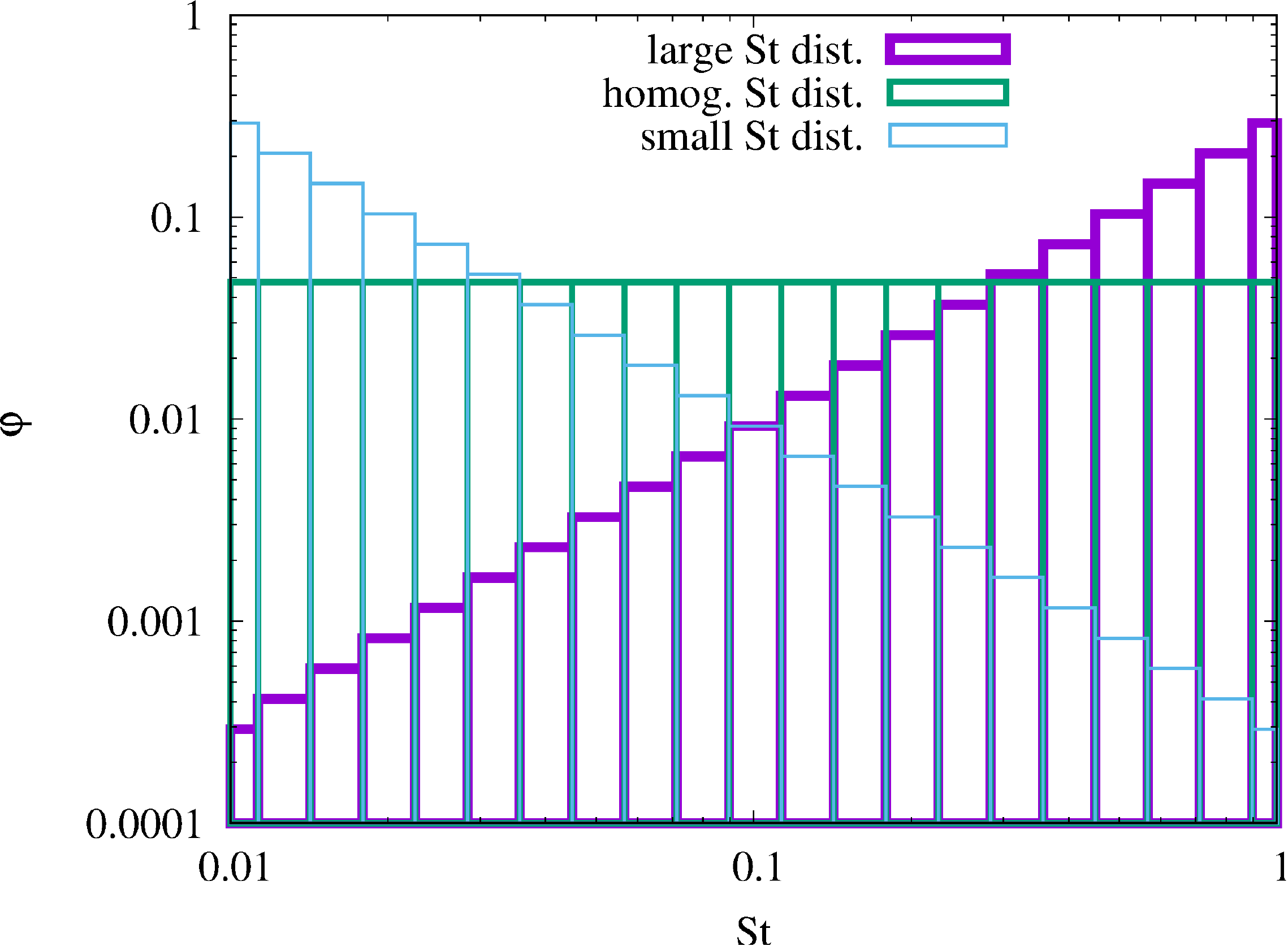} 
  \caption{Histograms of the mass-weights $\varphi_i$ corresponding to the species {\it i} for each of the three mass distributions considered in \S~\ref{sec:stokes_number_distri}.} 
    \label{fig0_sec3.4}
\end{figure}

We briefly recall that for a mass distribution with $\text{d}n/\text{d}m \propto m^{-\beta}$, the total dust mass is $\propto m^{2-\beta}$. Thus, depending on whether $\beta<2$ or $\beta>2$, most of the mass of the system is contained either in the larger or the smaller particles, respectively. In addition, it is easy to show that the size distribution is given by $\text{d}n/\text{d}a \propto a^{3\beta-2}$. In the Epstein drag regime, the Stokes number is proportional to the particle size and thus $\text{d}n/\text{d}\text{St} \propto \text{St}^{3\beta-2}$. 

We consider three cases covering different characteristic possibilities: a distribution for which most of the mass is in small particles, with $\beta=5/2$ and $\text{d}n/\text{d}\text{St} \propto \text{St}^{-5.5}$, a homogeneous distribution, with $\beta=2$ and  $\text{d}n/\text{d}\text{St} \propto \text{St}$, and a distribution for which most of the mass is in larger particles, with $\beta=3/2$ and $\text{d}n/\text{d}\text{St} \propto \text{St}^{-2.5}$. For context, in the classical fragmentation equilibrium case $\beta=11/6$ and thus $\text{d}n/\text{d}\text{St} \propto \text{d}n/\text{d}a \propto a^{-3.5}$  \citep{Dohnanyi69}, implying particle distributions for which most of the mass is in larger particles.

We represent the distribution of Stokes numbers with 21 evenly spaced bins in logarithmic-scale between St= 0.01 and St= 1. The dust species $i$ contributes with a dust-to-gas mass ratio $\epsilon_i = \varphi_i \epsilon$, such that $\sum_i \varphi_i= 1$ and thus $\sum_i \epsilon_i= \epsilon$.
In Fig.~\ref{fig0_sec3.4} we show the histograms of the mass weights for the three distributions considered. Note that in the case of $\text{d}n/\text{d}m \propto m^{-5/2}$ the four smallest Stokes numbers in the distribution contribute with 75\% of the mass, whereas for $\text{d}n/\text{d}m \propto m^{-3/2}$ the four largest Stokes numbers in the distribution contribute 75\% of the mass. In the homogeneous distribution each species has the same $\varphi_i= 1/21$. 

The total dust torque can be computed as the sum of the dust torque exerted by each individual dust species\footnote{This is valid provided the dust does not transfer momentum to the gas and all dust species interact only via the gas.}
\begin{equation}
\Gamma_{\text{d}}= \sum_{i}\Gamma_{\text{d}}^{i},  
\label{eq1_sec3.4}
 \end{equation} 
where $\Gamma_{\text{d}}^{i}$ can be obtained from Eq.~(\ref{eq:dust_torque}) by setting $\epsilon = \epsilon_i$ for each species.

\begin{figure}[t!]
    \centering
    \includegraphics[width=1.\columnwidth]{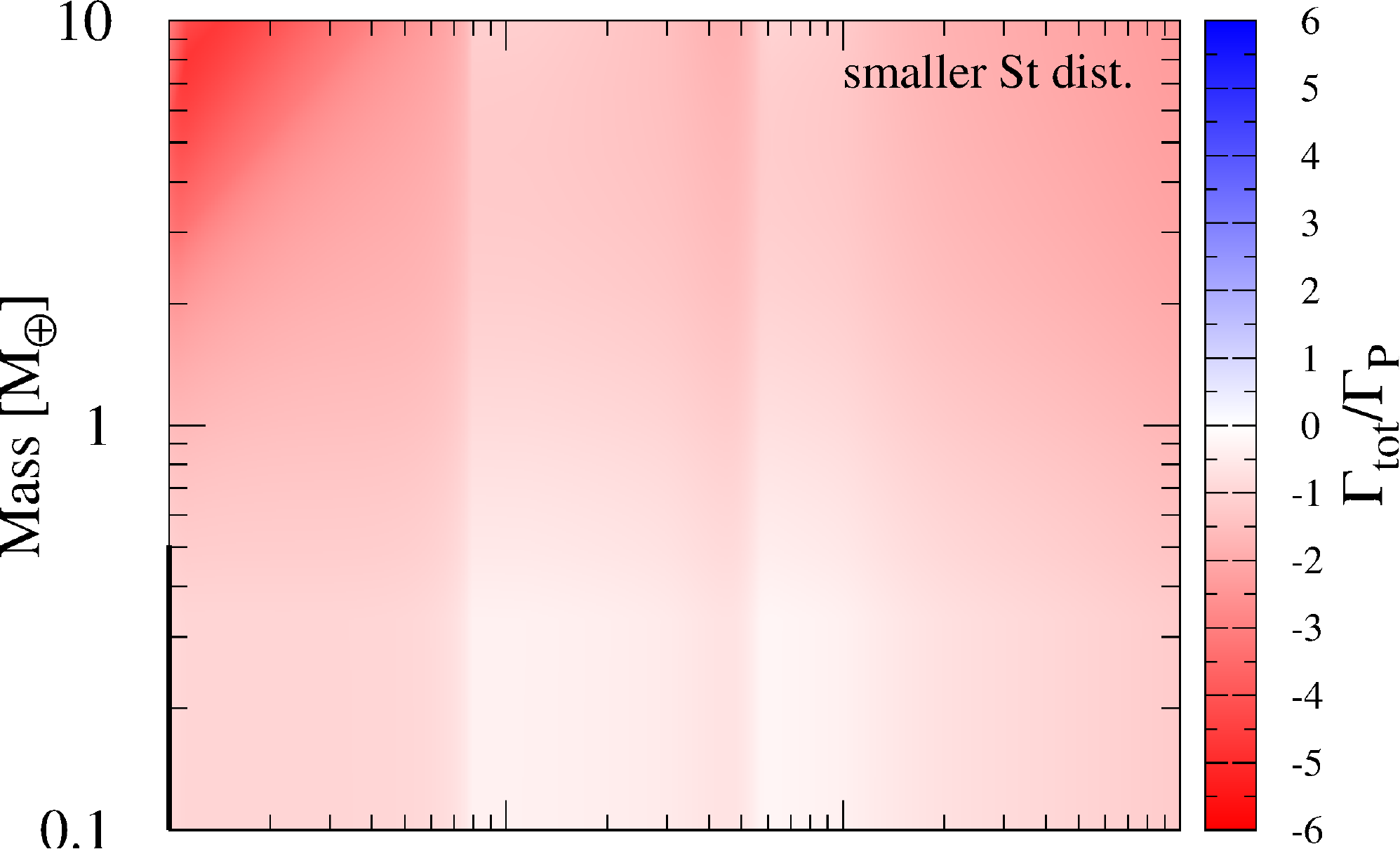} 
    \\
    \centering
    \includegraphics[width=1.\columnwidth]{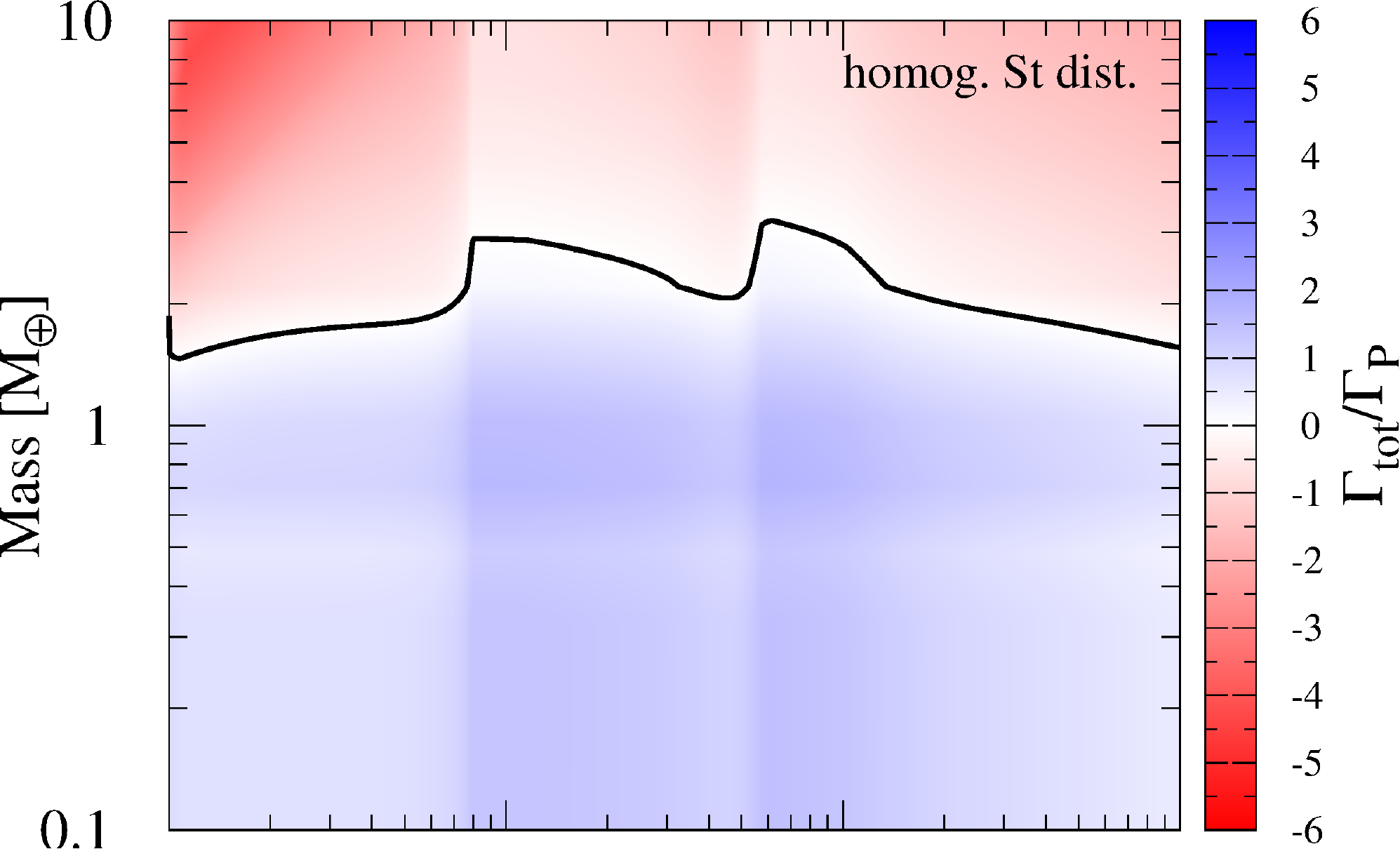} \\
    \includegraphics[width=1.\columnwidth]{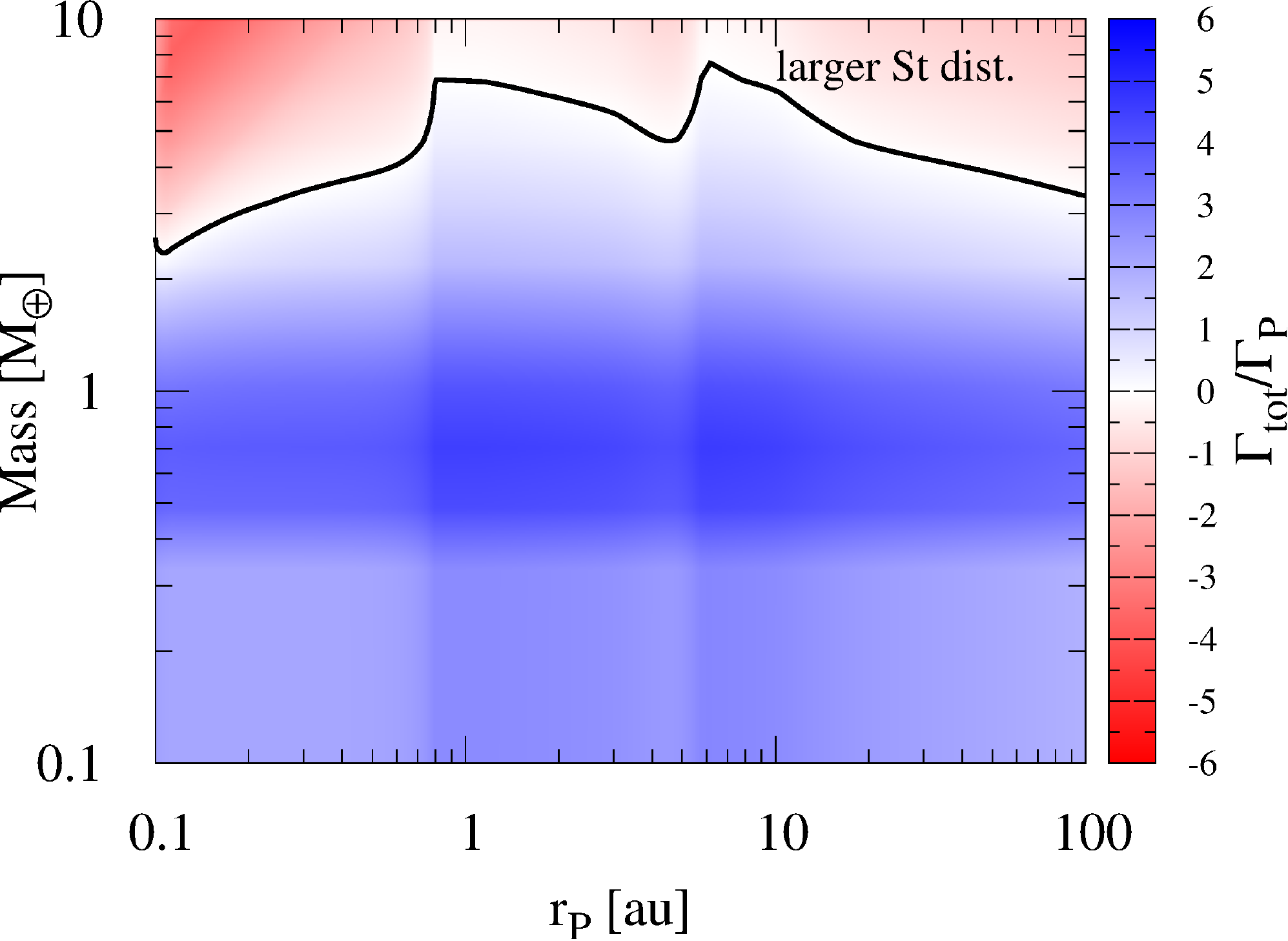}
    \caption{Same as Fig.~\ref{fig:torque_map_mp_vs_rp_non_iso} but considering a Stokes number distribution between St= 0.01 and St= 1. The top panel shows the case where particles with the smaller size/Stokes number contribute the most to the mass of solids in the disk. The middle panel represents the case of a homogeneous mass (or Stokes number) distribution, while the bottom panel corresponds to the case where the mass of solids in the disk is dominated by the larger particle-size/Stokes numbers.} 
    \label{fig1_sec3.4}
\end{figure}

The torque maps obtained for each distribution are shown in Fig.~\ref{fig1_sec3.4}. In the case where most of the mass is in the particles with lower Stokes numbers, the torque map is similar to the gas torque (left panel in Fig.~\ref{fig:torque_map_mp_vs_rp_non_iso}) and the solids do not play a significant role. On the other hand, for the homogeneous distribution and for the distribution where most of the mass is contained in large Stokes number particles (middle and lower panels in Fig.~\ref{fig1_sec3.4}) the dust torque dominates and generates regions of positive torque for planets with masses up to $\simeq 3~\text{M}_{\oplus}$ and $\simeq 7~\text{M}_{\oplus}$, respectively. These results suggest that dust-growth models in which most of the mass remains in the form of large particles \citep[e.g.,][]{Birnstiel12} favor outward migration of low-mass planets provided these particles have $\text{St} \simeq 0.1$--$1$. As we show in \citetalias{Guilera22b}, these results may have important implications for the role of the dust torque on the migration of low-mass planets growing by pebble accretion.

\section{Planet migration} 
\label{sec:planet_migration}

Using the models for the gas and dust torques as described in \S\ref{sec:TorqueCalculation}, we can integrate the equations of motion for an embedded planet to obtain its migration history in the disk. In order to accomplish this, we implement a simple model in which the planet mass grows by accretion of pebbles with a fixed Stokes number. For simplicity and consistency, we evolve only one planet at a time. We consider initial conditions at $r_{\rm p} =\{1, 5, 10\}$~au and the initial mass of the planet is $0.1~\text{M}_{\oplus}$ in all cases. The orbital integration is terminated when any of the following conditions is met: the planet grows to 10~$\text{M}_{\oplus}$, or the planet reaches the inner/outer disk edge located at 0.1/100~au. We note here that in general these conditions are fulfilled in less than $\sim 1$~Myr for the cases of moderate and high pebble accretion rates ($10^{-5}~\text{M}_\oplus/\text{yr}$ and $10^{-4}~\text{M}_\oplus/\text{yr}$, respectively) and in less than 5~Myr for the case of a low pebble accretion rate ($10^{-6}~\text{M}_\oplus/\text{yr}$).

\subsection{Steady-state Isothermal Disks Models} 
\label{planet_migration_isothermal}

\begin{figure*}[t!]
  \centering
  \includegraphics[width=\textwidth]{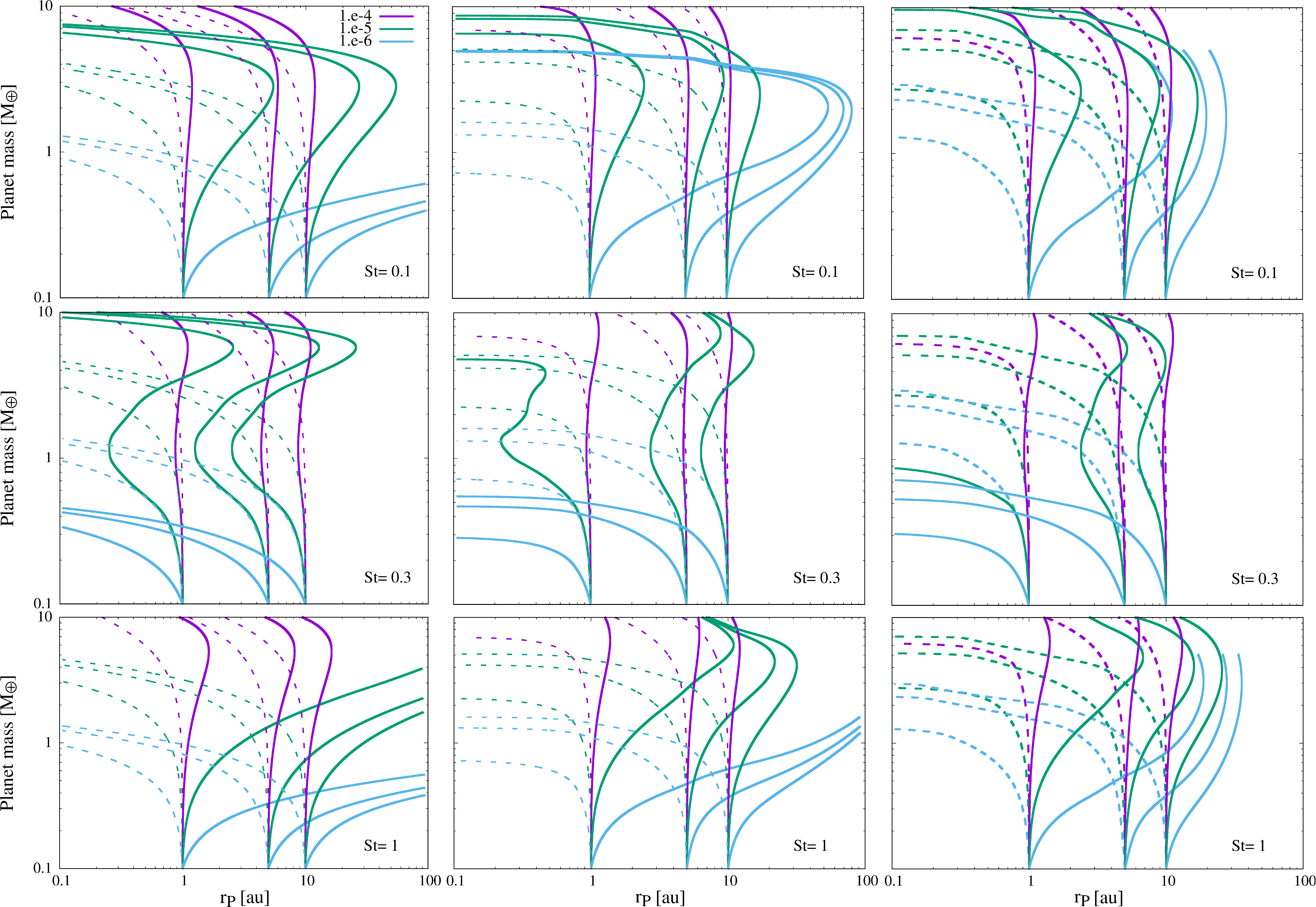} 
  \caption{Formation tracks for planets initially located at $r_{\rm p} =\{1, 5, 10 \}$~au evolving in different disk models.  The results corresponding to  steady-state isothermal disks (see \S~\ref{sec:FiducialModel}), steady-state non-isothermal disks (see \S\ref{sec:non-isothermal disk}), and time-dependent non-isothermal disks (see \S\ref{planet_migration_non-isothermal_evolving}) are shown in the left, center, and right set of panels.  Each row of panels shows a given Stokes number in St$=\{0.1, 0.3, 1\}$. Solid lines represent tracks of planets in dusty disks, with a constant dust-to-gas mass ratio $\epsilon = 0.01$, growing via pebble accretion with $\dot M_{\rm p} = \{10^{-4}, 10^{-5}, 10^{-6}\}~\text{M}_\oplus/\text{yr}$ (violet, green, and cyan lines respectively). Dashed lines show the corresponding formation tracks in dust-free disks.} 
  \label{fig1_sec4}
\end{figure*}

We first consider a steady-state, vertically isothermal $\alpha$-disk model with $\alpha=3\times10^{-3}$ and constant stellar accretion rate $\dot{M}_{\star}= 10^{-7}~\text{M}_{\odot}/\text{yr}$. When dust is present, the dust-to-gas mass ratio is $\epsilon = 0.01$ and the Stokes number takes a fixed value with St$=\{0.1, 0.3, 1\}$. We consider three constant values for the accretion rate of solids onto the planet according to $\dot M_{\rm p} = \{10^{-4}, 10^{-5}, 10^{-6}\}~\text{M}_\oplus/\text{yr}$. The resulting formation tracks in dusty disks are shown in the left column in Fig.~\ref{fig1_sec4} with solid lines. For reference, the formation tracks in the dust-free disks, exhibiting inward migration, are shown with dashed lines. We remark here that, under our steady-state assumption, the high mass accretion rate assumed implies that the disk gas surface density remains high during the planet's formation and migration. In this regard, this choice provides a reference case for the early stages of a more realistic disk evolution model. We note that, in steady-state, the associated integrated disk mass between 0.1~au and 40~au is about $0.1~\text{M}_{\odot}$, which corresponds to a gravitational stable disk under the classical Toomre's stability criterion. The total gaseous disk mass between 0.1~au and 100~au is $\sim 0.3~\text{M}_{\odot}$, which is too massive for typical protoplanetary disks.  However, when we consider disk models that evolve in time below (\S~\ref{planet_migration_non-isothermal_evolving}), we implicitly assume that the disk is initially massive and compact and that, as the mass accretion rate decreases, the disk mass decreases and it spreads viscously.

As seen in Fig.~\ref{fig1_sec4}, planets migrate initially outwards for St= 0.1 and St= 1. In the case of St= 0.1 the outward migration continues until they reach $\simeq 3~\text{M}_\oplus$ (for $\dot M_{\rm p}=10^{-4}$ and $10^{-5}~\text{M}_{\oplus}/\text{yr}$) or until they reach 100~au (for $\dot M_{\rm p}=10^{-6}~\text{M}_{\oplus}/\text{yr}$). Similar behaviour occurs for St= 1, where planets migrate outwards until they reach $\simeq 5.5~\text{M}_\oplus$ for $\dot M_{\rm p}=10^{-4}~\text{M}_{\oplus}/\text{yr}$, while for $\dot M_{\rm p}=10^{-5}$ and $10^{-6}~\text{M}_{\oplus}/\text{yr}$ they continue migrating outwards until they reach 100~au. 

For St= 0.3, planets migrate inwards at early times in all the cases, in agreement with the torque map showed in Fig.~\ref{fig:torque_map_mp_vs_rp_iso}. For the case of a solid accretion rate of $\dot M_{\rm p}=10^{-6}~\text{M}_{\oplus}/\text{yr}$, planets do not reach the needed mass ($\simeq 1~\text{M}_\oplus$) to revert the inward migration and they reach the inner disk at 0.1~au very fast with masses under $0.5~\text{M}_\oplus$. For $\dot M_{\rm p} = 10^{-5}~\text{M}_{\oplus}/\text{yr}$, planets revert the inward migration and migrate outwards until they reach $\simeq 6~\text{M}_\oplus$. After reaching this mass their migration reverses again so that they migrate fast inwards reaching $10~\text{M}_{\oplus}$ at the inner disk edge. For the highest solid accretion rate we consider, $\dot M_{\rm p}=10^{-4}~\text{M}_{\oplus}/\text{yr}$, the planets grow fast enough so that they do not migrate significantly before they reach $10~\text{M}_{\oplus}$.

Whether or not dust is present has a more significant impact on the formation tracks when the solid accretion rate is lower. Naturally, planets that reach $M_{\rm p}\simeq 10~\text{M}_{\oplus}$ fast enough do not have the chance to spend significant time in the region of the torque maps where the dust contribution is important. The situation is different when the accretion of solids is slower. In particular, for $\dot M_{\rm p} = 10^{-5}$ or $10^{-6}~\text{M}_{\oplus}/\text{yr}$, and dust with St= 0.1 and 1, planets can remain in disk regions where they are subject to positive torques for longer times, affecting their formation tracks with respect to the dust-free case significantly.

\subsection{Steady-state Non-isothermal Disks Models} 
\label{planet_migration_non-isothermal}

We now consider a steady-state, non-isothermal $\alpha$-disk model with $\alpha=3\times10^{-3}$ and constant stellar accretion rate $\dot{M}_{\star}= 10^{-7}~\text{M}_{\odot}/\text{yr}$ and evolve an embedded planet sweeping the same set of initial conditions and physical parameters we used for the vertically-isothermal disk. The formation tracks in steady-state, non-isothermal dusty disks are shown in the middle column in Fig.~\ref{fig1_sec4} with solid lines. For reference, the formation tracks in the dust-free disks, exhibiting inward migration, are also shown with dashed lines.

Overall, the planet formation tracks for non-isothermal disks are qualitatively similar to the ones computed for vertically-isothermal disks. 
Some of the results worth highlighting are: {\it i)} In dusty non-isothermal disks, in contrast to the isothermal case,  the transition from outward to inward (or inward to outward) planet migration does not occur at the same planet mass, in agreement with Fig.~\ref{fig:torque_map_mp_vs_rp_non_iso}. {\it ii)} In dusty, non-isothermal disks with St= 0.1 none of the planets reach 100~au when the solid accretion rate is $\dot M_{\rm p}=10^{-6}~\text{M}_{\oplus}/\text{yr}$. These planets reach $\simeq 2~\text{M}_{\oplus}$ and reverse their initial outward migration somewhere between $\sim 50$~au and $\sim 80$~au, depending on their initial locations. They all reach the inner disk edge with roughly the same mass $\sim 5~\text{M}_{\oplus}$. A similar behavior is observed for the case of St= 1 and $\dot M_{\rm p}=10^{-5}~\text{M}_{\oplus}/\text{yr}$. {\it iii)} Similarly to the isothermal disks case, for high rates of accretion of solids, e.g., $\dot M_{\rm p}=10^{-4}~\text{M}_{\oplus}/\text{yr}$, the planet formation tracks are rather insensitive to the presence of dust. This is especially true for the planets initially located at 5~au and 10~au, due to fast planet formation timescales. 

\subsection{Evolving Non-isothermal Disks Models} 
\label{planet_migration_non-isothermal_evolving}

We consider here a simple model for the temporal evolution of a non-isothermal disk as the stellar accretion rate decreases in time. We adopt the same model considered in \citet{Bitsch15a} in which the disk evolves through a succession of steady-state solutions wherein the mass accretion rate $\dot{M}_{\star}$ decreases in time according to 
\begin{equation}
\log \left( \dfrac{\dot{M}_{\star}}{\text{M}_\odot~\text{yr}} \right)= -8 - 1.4 \times \log \left( \dfrac{t + 10^5~\text{yr}}{10^6~\text{yr}} \right).
\label{eq_disk_evol}
\end{equation}
This parameterizes a smooth transition between a mass accretion rate of $ 10^{-7}~\text{M}_{\odot}/\text{yr}$ and $10^{-9}~\text{M}_{\odot}/\text{yr}$ in 5~Myr. As time advances and the mass accretion rate decreases, the gas surface density decreases accordingly. As in the steady-state disk models considered above, we set $\alpha= 3 \times 10^{-3}$. We compute a dense grid of 50 steady-state $\alpha$-disk models with equally-spaced accretion rates $\dot{M}_{\star}$ between the initial and final accretion rates. We interpolate between these 50 models as needed to obtain a smooth disk evolution in time. For simplicity, we assume that the dust-to-gas mass ratio $\epsilon$ remains constant as the disk density evolves in time. This provides a conservative approach to assess the impact of dust in altering the planetary formation tracks with respect to the dust-free disks.

In the background of these time-dependent, non-isothermal disk models, we evolve an embedded planet sweeping the same set of initial conditions and physical parameters we used for the steady-state disk models. The resulting formation tracks in dusty disks are shown in the right column in Fig.~\ref{fig1_sec4} with solid lines. For reference, the formation tracks in the dust-free disks, exhibiting inward migration, are shown with dashed lines.
Note that for the highest pebble accretion rate considered, $\dot M_{\rm p}=10^{-4}~\text{M}_{\oplus}/\text{yr}$, the formation tracks are similar to the ones in steady-state disks, irrespective of the Stokes number. This is due to the fact that formation timescales are shorter than migration timescales and planets form fast enough that the disk evolution does not affect the planet formation process. 

\begin{figure*}[t!]
  \centering
  \includegraphics[width=\textwidth]{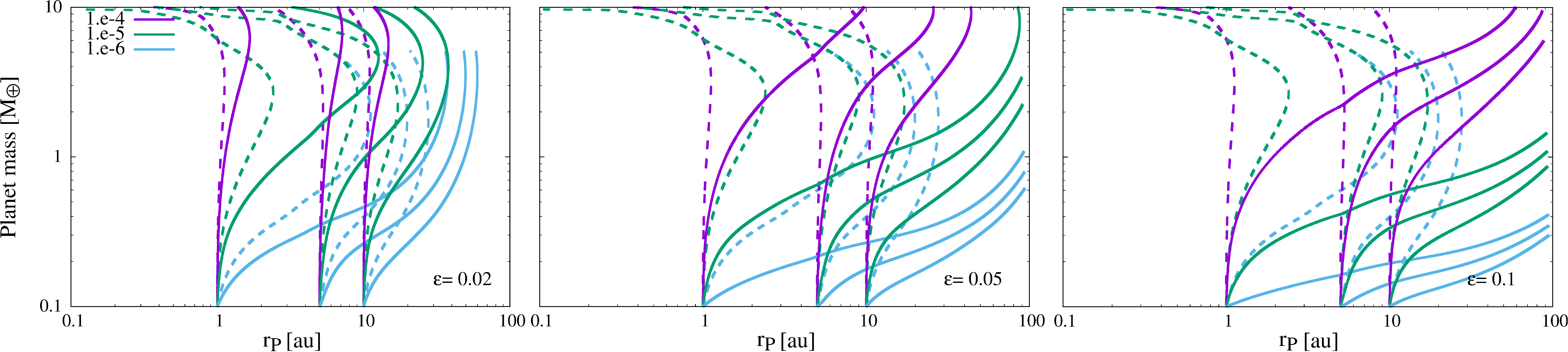} 
  \caption{Formation tracks for planets initially located at $r_{\rm p} =\{1, 5, 10 \}$~au evolving in a time-dependent non-isothermal disk (see \S\ref{planet_migration_non-isothermal_evolving}) with a solid component with Stokes number St = 0.1. The results corresponding to different gas-to-dust mass ratios in $\epsilon = \{0,02, 0.05, 0.1\}$ are shown in the left, center, and right panels, respectively. Solid lines represent tracks of planets growing via pebble accretion with $\dot M_{\rm p} = \{10^{-4}, 10^{-5}, 10^{-6}\}~\text{M}_\oplus/\text{yr}$ (violet, green, and cyan lines respectively). Dashed lines show the corresponding formation tracks in disks with $\epsilon = 0.01$.}
  \label{fig:eps_increase}
\end{figure*}

For the cases with pebble accretion rates $\dot M_{\rm p}=10^{-5}~\text{M}_{\oplus}/\text{yr}$, the planet formation tracks present some differences with respect to the steady-state, non-isothermal dust-free disk. However, these are not as noticeable as for the solid accretion rate $\dot M_{\rm p}=10^{-6}~\text{M}_{\oplus}/\text{yr}$. This is particularly the case for Stokes numbers St= 0.1 and St= 1, for which planets in steady-state disks end at 0.1~au with $5~\text{M}_{\oplus}$ and at 100~au with $2~\text{M}_{\oplus}$, respectively. In contrast, in evolving disks, all the planets reach $5~\text{M}_{\oplus}$ in 5~Myr in moderate-to-extended orbits, i.e., between $\simeq 5$~au and $\simeq 30$~au, depending on their initial locations. As the gas torque on the planet decreases, due to the decreasing gas surface density of the disk, the planets are allowed to remain in orbits with moderate to extended radii for longer times. Clearly, disk evolution plays an increasingly significant role the slower the planets grow.

\subsection{Impact of the Dust-to-gas Mass Ratio}
\label{planet_migration_dust-to-gas}

As stated in \S\ref{sec:torque_definitions}, in the absence of dust-feedback on the gas, the dust torque scales linearly with the dust-to-gas mass ratio. In order to explore the impact of the dust-to-gas mass ratio on the planet formation tracks we consider the set of values $\epsilon=\{0.02, 0.05, 0.1\}$. If protoplanetary disks have dust-to-gas mass ratio similar to the metallicity of their host protostars, values of up to $\epsilon= 0.03$ could be found in disks around metal-rich stars. Higher values of dust-to-gas mass ratios, in at least some disk regions, could be achieved via other mechanisms. For instance, the dust-to-gas mass ratio can increase in the inner disk regions due to dust drifting from larger radii \citep[e.g.,][]{Drazkowska2016}. Dust-to-gas mass ratios near unity have been envisioned to be reached in preferential locations in protoplanetary disks \citep[e.g.,][]{ Drazkowska2017}. As an example, we consider a fixed dust-to-gas mass ratio throughout the disk and compute the planet formation tracks for St= 0.1 assuming that the gaseous disk evolves in time according to Eq.~(\ref{eq_disk_evol}).

The impact of the dust-to-gas mass ratio on a sample of evolutionary tracks is shown in Fig.~\ref{fig:eps_increase}. It is seen that for dust-to-gas mass ratios with $\epsilon \gtrsim 0.05$ planets can migrate significantly outwards even for the highest solid accretion rate we considered, $\dot{M}_{\text{P}}= 10^{-4}~\text{M}/\text{yr}$. For the models with lower solid accretion rates, the effect of larger dust-to-gas mass ratios is increasingly dominant for outward planet migration, irrespective of Stokes number. As we show in a follow-up paper (\citetalias[][in preparation]{Guilera22b}), enhanced dust-to-gas mass ratios in the inner disk regions, e.g., due to dust drifting from outer disk regions, can have a significant impact on the migration of growing planetary cores inside the water-ice line.  

\section{Summary \& Discussion}
\label{sec:discussion}

While planet migration has been extensively studied in gaseous disks, the effect of solids on this process has not yet been addressed. Measurements of dust torques have been reported by BLP18, but the effect of these torques on the migration of planetary embryos has remained an open question. Aiming at filling this gap, in this work we quantify for the very first time the impact of the torque arising from solids on planet migration and assess its role on the formation tracks of planetary embryos embedded in classical protoplanetary disk models. 

Based on the torque measurements of BLP18, we computed what we refer to as ``torque maps'' in order to quantify the importance of the dust torque (with respect to the gas torque) in terms of the disk thermodynamics, the planet location, the stellar mass-accretion rate, the dust-to-gas mass ratio, the level of turbulent viscosity, and different dust-size distributions. Subsequently, we investigated how the migration of growing planets is affected by disk thermodynamics, dust content, and global evolution.

In order to make progress in an inherently complex problem, we have made a number of simplifying assumptions and approximations. This approach enabled us to obtain the first quantitative estimates for the impact of the dust torque under a very wide range of physical conditions.  Some of the effects we have found could become more or less prominent when models alleviating the caveats mentioned below are considered. Because of this, the theoretical formation path that a planetary core follows may depend on the details of the models employed.  Nevertheless, it is worth stating some of the consequences associated with the broad trends that emerge from the results we obtained using standard assumptions to model the disk and its solid content.

\subsection{Implications}

Our findings have a series of implications for the processes involved in the formation and evolution of planetary cores in dusty disks. These, in turn, determine not only the formation timescales, final masses, and location of the planetary embryos but, potentially, also the final compositions of the resulting planets. 
\paragraph{Low-mass planets migrate outwards beyond the water ice-line} Dust growth models suggest that most of the mass in solids is contained in particles in the high-end of the particle size distribution \citep[e.g.,][]{Birnstiel12, Stammler2022}. In addition, \citet{Venturini20Letter} and \citet{Drazkowska21} showed that the mass-averaged Stokes number of the dust distribution is St $\sim 0.1$ beyond the water ice-line (see their Fig. 1 and Fig. 4, respectively). Thus, the results presented in  \S~\ref{sec:stokes_number_distri} imply that forming planets of up to a few Earth-masses located beyond the water ice-line migrate outwards, unless the dust-to-gas mass ratio becomes lower than $\sim 10^{-3}$ (see \S~\ref{sec:dust_to_gas_ratio}).

\paragraph{Small-Stokes numbers could matter in dust-rich disks} Under a number of circumstances, the dust-to-gas mass ratio could increase significantly in some disk regions. This could occur in the inner part of the disk due to radial pebble drift from the outer disk \citep[e.g.,][]{Drazkowska2016, Venturini20ST}, near the water ice-line due to water vapor re-condensation \citep[e.g.,][]{Drazkowska2017,Schneider21}, or in a pressure bump where pebbles could accumulate  \citep[e.g.,][]{Pinilla2012, Zhu2012, Weber2018, Morby2020, Guilera20,Chambers2021, Jiang2023}.  This implies that if the dust-to-gas mass ratio reaches, for example, $\epsilon = 0.1$, even particles with Stokes numbers as small as St= 0.01 can exert significant torques if most of the mass is in those particles (Fig.~\ref{fig1_sec4.1}).  

\paragraph{Forming planets in extended orbits in dust-rich disks} Disks with large content of solids favor outward migration. This may offer a mechanism for the cores that start their formation in the inner disk region to end up located far from the central star, at tens of au (see Fig.~\ref{fig1_sec4} and Fig.~\ref{fig:eps_increase}). Moreover, these planets that migrate outwards, up to tens of au, could act as seeds for the formation of more massive planets in wide orbits. This could offer a natural path to explain the formation of these controversial objects, which are usually invoked as responsible for the ring and gap structures observed in protoplanetary disks by ALMA \citep[e.g.,][]{Andrews2020}. 

\subsection{Current Caveats and Future Lines of Work}

Here, we state and assess our most important hypothesis and mention future lines of work aimed at building more realistic models. More accurate and consistent models for the gas (and dust) torque could be obtained with dedicated, systematic numerical simulations considering more detailed disk models. 

\paragraph{Precomputed prescriptions for type-I migration} Even though the disk models we consider have a non-trivial radial and thermodynamic structure, we adopted the torque prescriptions of \citet{Tanaka02}, derived for globally isothermal disks with power-law radial structure, and \citet{jm2017}, derived for adiabatic disks. These prescriptions for the gas torque neglect migration feedback \citep[see, e.g.,][]{Paardekooper2014}, an effect that is more pronounced in low-viscosity disks \citep[see the review of][]{Paardekooper2022}.

\paragraph{Caveats inherited from BLP18} By construction, our results contain all the caveats involved in computing the dust torques in BLP18. In particular, the numerical simulations involved are two-dimensional and thus insensitive to the vertical disk structure. More realistically, in three dimensions, the results are expected to depend on the vertical distribution of the solid component, which is sensitive to the level of vertical diffusion of solids, as regulated by the level of turbulence in the disk. Additional studies of dust torques properly considering the turbulent diffusion of solids are needed to better quantify the magnitude of dust torques. Moreover, simulations in BLP18 only considered disks with radial $(-1/2)$ power-law gas-surface density structures. How the magnitude of dust torques is affected by the underlying gas-disk model is still uncertain, and could potentially affect some detailed aspects of our results. In particular, we assumed that dust torque does not depend on the gas torque (but see Appendix \S~\ref{sec_app1}), which implicitly assumes that the underlying gas-disk model is not significantly important. This, however, may not be necessarily the case; more specifically, for solids that are highly coupled to the gas. Also, in BLP18 it is assumed that solids behave as a pressure-less fluid, which is reasonable for solids that are very well coupled to the gas. However, when solids are barely coupled, this approximation could break down. Furthermore, the dust torque measurements of BLP18 do not include the effect of planetary motion. Migration modifies the relative drift between solids and the planet and is expected to modify the shape and magnitude of the dust-distribution asymmetry developed close to the planet, which could either enhance or diminish dust torques. Moreover, if the planet migrates, it may develop eccentricity. Additional hydrodynamical simulations are needed in order to quantify the impact of eccentricity on dust torques.
It is also worth recalling that the effective smoothing length used in the two-dimensional simulations in BLP2018 can affect the measured values of dust torques, as confirmed by \cite{Regaly2020}. The dust torque in the gravity-dominated regime is robust because it arises from a large-scale asymmetry. However, in the gas-dominated regime the dust torque occurs due to a subtle asymmetry in the dust density distribution close to the planet (BLP2018). In order to minimize spurious measurements due to unresolved dynamics, BLP18 conservatively cut off the inner half of the Hill sphere to compute the torques. Nevertheless, three-dimensional simulations, where the smoothing length is not a free parameter, are needed to quantify more accurately the dust torque in different regimes.
Thermodynamic effects could be important for dust torques. For example, the equation of state used could affect the gas dynamics and may have a non-negligible effect on the solid dynamics \citep[see, e.g.,][]{Miranda2019}. Also, the effect of pebble accretion can change the dust asymmetry around the planet and hence the magnitude of the dust torque \citep[see, e.g.,][]{Regaly2020}, and generate heat that would lead to heating torques \citep[e.g.,][]{Benitez-llambay2015, masset2017}, an effect that, for simplicity, we do not consider in this first work. Finally, the dust back-reaction force (feedback) is neglected in BLP18. This implicitly assumes that the density contrast of solids remains low throughout the disk and close to the planet. However, large density contrast of solids is observed in the numerical simulations in BLP18, which may have an impact on the dynamics of gas and solids close to the planet. Further work including the dust feedback is needed to assess the impact of this approximation.

\paragraph{Extrapolating BLP18's results to different disk models} In order to extrapolate the results of BLP18 to other disk models we assumed a particular scaling of the dust torque with the disk parameters, see Eq.~(\ref{eq:dust_torque}). However, the dust torque could scale additionally with other physical disk parameters that we do not consider here. In order to assess the robustness of our results, in Appendix \ref{sec_app1} we assume an alternative model where the dust torque is proportional to the gas torque, see Eq.~(\ref{eq:dust_torque_app}), and compare the results. We show that our main results and conclusions are insensitive to the specific choice.

\paragraph{Pebble accretion, Stokes number/dust-size distributions} In order to compute the migration of the growing planetary embryos, we adopted constant pebble accretion rates and Stokes numbers (see \S~\ref{sec:planet_migration}). A number of models usually adopt solid particles with constant Stokes number -or constant size- throughout the disk with pebble fluxes that decay in time, decreasing the pebble accretion rate on the planet \citep[e.g.,][]{Lambrechts19, Ogihara2020}. In more realistic planet formation models that include dust growth and evolution, both the pebble Stokes numbers and the pebble flux evolve in time, and throughout the disk, in a more complex way \citep[e.g.,][]{Venturini20Letter, Drazkowska21}. 

\paragraph{Isolated planets} Our study assumes that a planet evolves in the disk in isolation. In order to extrapolate our results to disks containing multiple planets it would need to be assumed that migrating planets can be modeled independently of each other. This assumption will not hold in compact systems where the gravitational interaction between cores and planets could be significant. Also, multiple cores migrating are expected to exert strong perturbations on the dust surface density that can propagate from the outermost planet inwards via radial drift (see, e.g., Fig.~4 in BLP2018). The conditions for significant interference between multiple planets could in principle be assessed with targeted numerical simulations. In addition, when multiple planets grow simultaneously if one of them becomes massive enough to open a gap, or a partial gap, on the disk, the flux of pebbles to the planets in inner orbits can be halted \citep[e.g.,][]{Lambrechts14, Weber2018}. This situation could also affect the magnitude of the dust torque for the inner planets.  

\paragraph{Disk gaps induced by photoevaporation}
The classical disk models we adopted neglected, for simplicity,  the effects of photoevaporation due to the central star. In low-mass stars, X-rays are the dominant source of photoevaporation \citep[see, e.g.,][]{Kunitomo2021}. \citet{Venturini20ST} showed that X-ray photoevaporation can open a gap in the gas disk at a few au on a typical timescale of 1-2~Myr. This gap prevents the pebble flux from reaching the inner disk, significantly affecting pebble accretion onto the planets there. Thus, more realistic gas disk evolution models should also affect the computation of the dust torque. In a follow-up paper (\citetalias[][in preparation]{Guilera22b}), we incorporate the computation of the dust torque in {\scriptsize PLANETALP}, our global framework for modeling planet formation, in order to study its effects on planet migration in more realistic settings for modeling planet formation. 

\paragraph{Disk viscosity} Our work is based on the assumption that the disk remains laminar so we can describe the gaseous torques by simple prescriptions. Also, we used measurements of dust torques in BLP18 in laminar viscous disks. However, planets embedded in low-viscosity disks can induce vortices, which affect not only the gas \citep[see e.g.][]{Paardekooper2014,McNally2019}, but also the dust torque. Thus, caution is needed when extrapolating our results to low-viscosity or wind-driven disks.

\section{Takeaways and Conclusion}
\label{sec:takeways}

Here, we summarize the most relevant takeaways from our study emphasizing the results that we expect not to depend too sensitively on our main assumptions and approximations.

\subsection{Main Takeaways}

- The dust torque can contribute significantly to the total torque acting on a planetary embryo, $M_{\rm p} \lesssim 10 M_\oplus$, for standard values of the dust-to-gas mass ratio, i.e., $\epsilon =0.01$ (see Figs.  \ref{fig2_sec2.4}, \ref{fig3_sec3}, and \ref{fig:torque_map_mp_vs_rp_non_iso} ).

- The dust torque is positive in the vast majority of the parameter space spanned by $0.1 \leq M_{\rm p}/M_\oplus \leq 10$ and $0.01 \leq \text{St} \leq 1$ for $\epsilon = 0.01$. The relative magnitude between the gas and dust torque components depends on the Stokes numbers of dust particles. 

- There are two regimes where the dust torque dominates significantly for $M_{\rm p} \lesssim 10 M_\oplus$; a ``gas-dominated'', $\text{St} \simeq 0.1$, and a ``gravity-dominated'' regime, $\text{St} \gtrsim 0.5$. These regions are associated, locally, with outward migration. This outcome is robust for a wide range of disk parameters such as the stellar mass accretion rate, the $\alpha$-viscosity parameter, and/or the disk thermodynamics adopted.

- In the transition region between the previous two, i.e., $\text{St} \simeq$ 0.2\textendash 0.5 and $M_{\rm p}/M_\oplus \leq 10$, the value of the dust torque is sensitive to all the parameters involved (e.g., viscosity, stellar-accretion rate, the planet location) and the direction of migration depends on details.

- If there are disk regions with enhanced dust-to-gas mass ratios of order $\epsilon = 0.1$, even particles with Stokes numbers as low as $\text{St} \lesssim 0.01$ can exert a significant torque (see Fig. \ref{fig1_sec4.1}). At these higher dust concentrations, the torque exerted by larger dust particles can influence the dynamics of planets with masses larger than $M_{\rm p} \simeq 10 M_\oplus$.

- The results outlined above do not depend sensitively on the stellar mass accretion rate or the $\alpha$-viscosity parameter (\S~\ref{sec:dependence_mass_accretion_rate} and \S~\ref{sec:dependence_viscosity}, respectively) or whether the dust particles are assumed to have a unique Stokes number or are distributed in a range.

- The properties of the torques as described above dictate the local dynamics of a planetary embryo in a dusty disk. Given that the relative magnitude of the gas and dust torque components evolves as the embryo migrates and grows, its long-term dynamics depends on the accretion rate onto the planet, the global disk structure and/or its long-term evolution.  

- Even though the details of the evolutionary tracks are sensitive to various physical parameters, we find that under the set of assumptions usually invoked dust torques can have a significant impact, during significant fractions of the formation history of planetary embryos, by halting or reversing inward migration. The impact of the dust component naturally decreases for planetary embryos that grow too fast, e.g., $\dot M_{\rm p} \simeq 10^{-4}~\text{M}_{\oplus}/\text{yr}$, or disks that run out of material too quickly (see Fig. \ref{fig1_sec4})

- The most important implications of our findings can be summarized as follows: 
{\it i)}
Low-mass planets migrate outwards beyond the water ice-line if most of the mass in solids is in larger particles there.
{\it ii)}
The torque due to solids with small Stokes numbers, St $\simeq 0.01$, can play a dominant role in disks with a moderately high dust-to-gas mass ratio, $\epsilon \simeq 0.02-0.05$ if most of the mass is in those particles.
{\it iii)} Under a wide range of conditions, dust torques could enable low-mass planetary cores formed in the inner disk to migrate outwards and act as the seed for massive planets at distances of tens of au, which could be
responsible for the large-scale structures observed  by ALMA.

\subsection{Conclusion}

Our most important conclusion is that, under a wide range of conditions that are usually assumed to hold when modeling planet formation processes, dust torques can have an important effect on the evolution of planetary embryos. This may well be an important fundamental piece missing in the puzzle of planet migration and formation \citep{2016JGRE..121.1962M}.

The detailed outcome when computing evolutionary tracks depends on the dust content and the global evolution of the disk. However, it is clear that the fast-inward migration of low-mass planetary embryos, known to affect gaseous disks, can be significantly alleviated as a natural consequence of the torques exerted by dust. As a result, the long-term orbital evolution of planetary embryos can differ substantially from those obtained in dust-free disks. Given the importance of the role of dust and planet migration in planet formation, we hope that our findings will motivate further studies of the processes involved to develop a more complete and coherent framework for making progress under more astrophysically realistic assumptions. \\


{\it We are grateful to the referee for a thoughtful report and useful suggestions that helped us improve the manuscript. OMG is partially supported by PICT 2018-0934 from ANPCyT, Argentina. OMG and M3B are partially supported by PIP-2971 from CONICET (Argentina) and by PICT 2020-03316 from Agencia I+D+i (Argentina). OMG acknowledges support by ANID, -- Millennium Science Initiative Program -- NCN19\_171. OMG and M3B also thank Juan Ignacio Rodriguez from IALP for the computation managing resources of the Grupo de Astrof\'{\i}sica Planetaria de La Plata. P.~B.~L. acknowledges support from FONDECYT project 1231205. MEP gratefully acknowledges support from the Independent Research Fund Denmark via grant no. DFF 8021-00400B and the Institute for Advanced Study, where part of this work was done.}

\appendix

\section{An Alternative Approach for Computing the Dust Torque}
\label{sec_app1}
An alternative way of computing the dust torque from the hydrodynamical simulations of BL18 consists of assuming that the dust torque $\Gamma_{\text{d}}$ is proportional to the gas torque $\Gamma_{\text{g}}$ instead of the reference torque $\Gamma_{\text{p}}$, as it is done in \S~\ref{sec:torque_definitions}. Under this assumption, the dust torque becomes
\begin{equation}
\frac{\Gamma_{\rm d}(r_{\rm p}, \epsilon)}{\Gamma_{\rm g}(r_{\rm p}, \epsilon)} = 
\left(\frac{\epsilon}{\epsilon_0}\right) \times \frac{\Gamma_{\rm d}(r_{\rm p}, \epsilon_0)}{\Gamma_{\rm g}(r_{\rm p}, \epsilon_0)} = 
\left(\frac{\epsilon}{\epsilon_0}\right) \times \frac{\Gamma_{\rm d}(r_{0}, \epsilon_0)}{\Gamma_{\text{g}}(r_{0}, \epsilon_0)}\,,
\end{equation}
and thus
\begin{eqnarray}
\Gamma_{\rm d}(r_{\rm p}, \epsilon) &=& \left(\frac{\epsilon}{\epsilon_0}\right) 
 \times
\frac{\Gamma_{\rm d}(r_{0}, \epsilon_0)} 
{\Gamma_{\text{g}}(r_{0}, \epsilon_0)}
\times
\Gamma_{\rm g}(r_{\rm p}, \epsilon) 
 = \left(\frac{\epsilon}{\epsilon_0}\right)  \times
 \left(\frac{{\Gamma_{\rm d}}}{\Gamma_{\text{g}}}\right)_{\text{BLP18}} \times \Gamma_{\rm g}(r_{\rm p}, \epsilon)
  = \Phi(M_{\rm p}, {\rm St}, \epsilon) \times \Gamma_\text{g}(r_{\rm p}, \epsilon)  \,,
\label{eq:dust_torque_app}
\end{eqnarray}
where $\Gamma_{\text{g}}(r_0, \epsilon_0)$ is the gas torque from BLP18 computed adopting $\epsilon_0=0.01$ and $r_0 = 1$ au\footnote{Note that the gas torque in BLP18 does not depend on the dust-to-gas mass ratio when this ratio is sufficiently small.}.
The ratio $(\Gamma_{\rm d}/\Gamma_{\text{g}})_{\text{BLP18}}$, as a function of planet mass, can be obtained\footnote{For the values of the planet masses ans Stokes numbers that are not tabulated we use bilinear interpolation to obtain the ratio $\Gamma_{\rm d}/\Gamma_{\text{g}}$.} from the information provided in Tab.~\ref{table:torques}. Finally, $\Gamma_\text{g}(r_{\rm p}, \epsilon)= \Gamma_\text{g}(r_{\rm p})$ is computed for the corresponding gas disk model as described in \S~\ref{gaseous-torque}.

\begin{figure*}[t!]
    \centering
    \includegraphics[width=1.\textwidth]{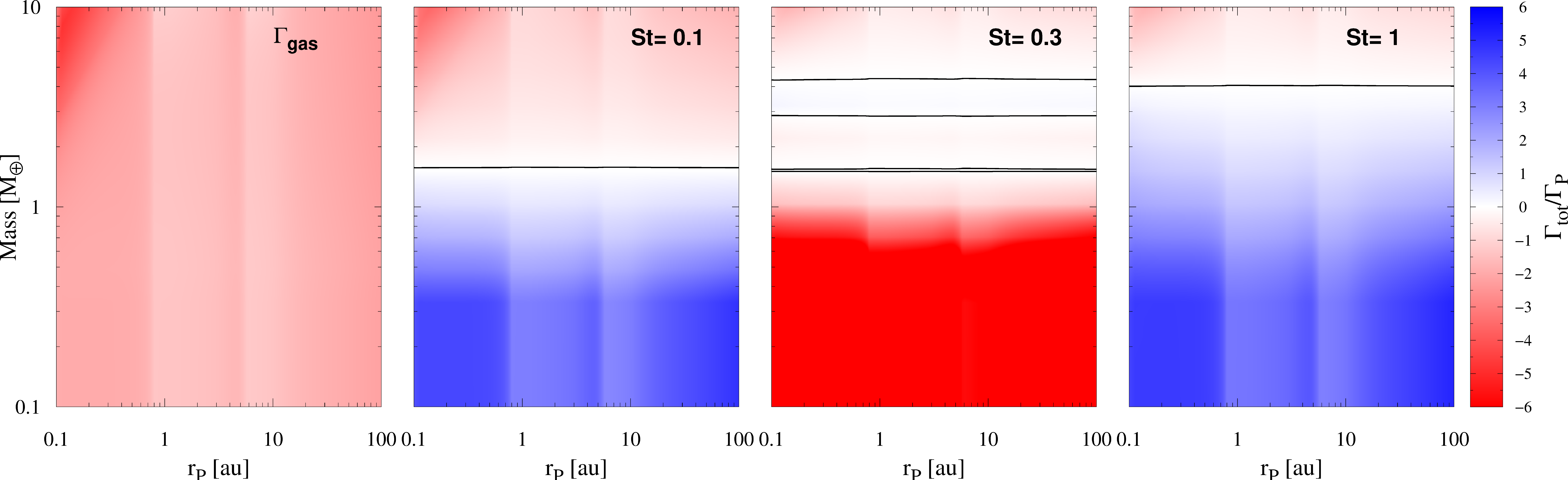} 
    \caption{Same as Fig.~\ref{fig:torque_map_mp_vs_rp_non_iso}, but here the dust torque is computed using Eq.~(\ref{eq:dust_torque_app}) instead of Eq.~(\ref{eq:dust_torque}).} 
    \label{fig1_app1}
\end{figure*}

In order to understand the impact of the model adopted for the dust torque, Fig.~\ref{fig1_app1} shows the normalized total torque for the case of a non-isothermal disk (see \S~\ref{sec:non-isothermal disk}) when the dust torque is assumed to be proportional to the gas torque instead of the reference torque $\Gamma_{{\text{p}}}$, as it was previously done using Eq.~(\ref{eq:dust_torque}).
As in Fig.~\ref{fig:torque_map_mp_vs_rp_non_iso}, the left panel in Fig.~\ref{fig1_app1} shows the total torque over a planet in a dust-free disk, i.e., $\Gamma_{\text{tot}}=\Gamma_{\text{g}}$. The following three panels show the total torque $\Gamma_{\text{tot}}=\Gamma_{\text{g}} + \Gamma_{\text{d}}$ for the Stokes numbers St= \{0.1, 0.3, 1\}, considered to be constant throughout the disk, and assuming a gas-to-dust mass ratio $\epsilon = 0.01$. 

The results shown in Fig.~\ref{fig1_app1} and Fig.~\ref{fig:torque_map_mp_vs_rp_non_iso} are qualitatively similar. The main difference is that in the latter the transitions between negative and positive torques in dusty disks occur at constant planet mass. This is because 
when the dust torque $\Gamma_{\text{d}}$ is proportional to the gas torque $\Gamma_{\text{g}}$ they both have the same radial dependence. The fact that this boundary is independent of the planet mass can be seen by noting that the zero torque condition $\Gamma_{\text{tot}}=0$ corresponds to $0= (1+\Phi(M_{\rm p}, {\rm St},\epsilon)) \times \Gamma_{\text{g}} $, which is satisfied by the same value of the planetary mass $M_{\rm p}$ regardless of the location of the planet, $r_{\text{p}}$. 

\begin{figure}[t!]
    \centering
    \includegraphics[width=.5\columnwidth]{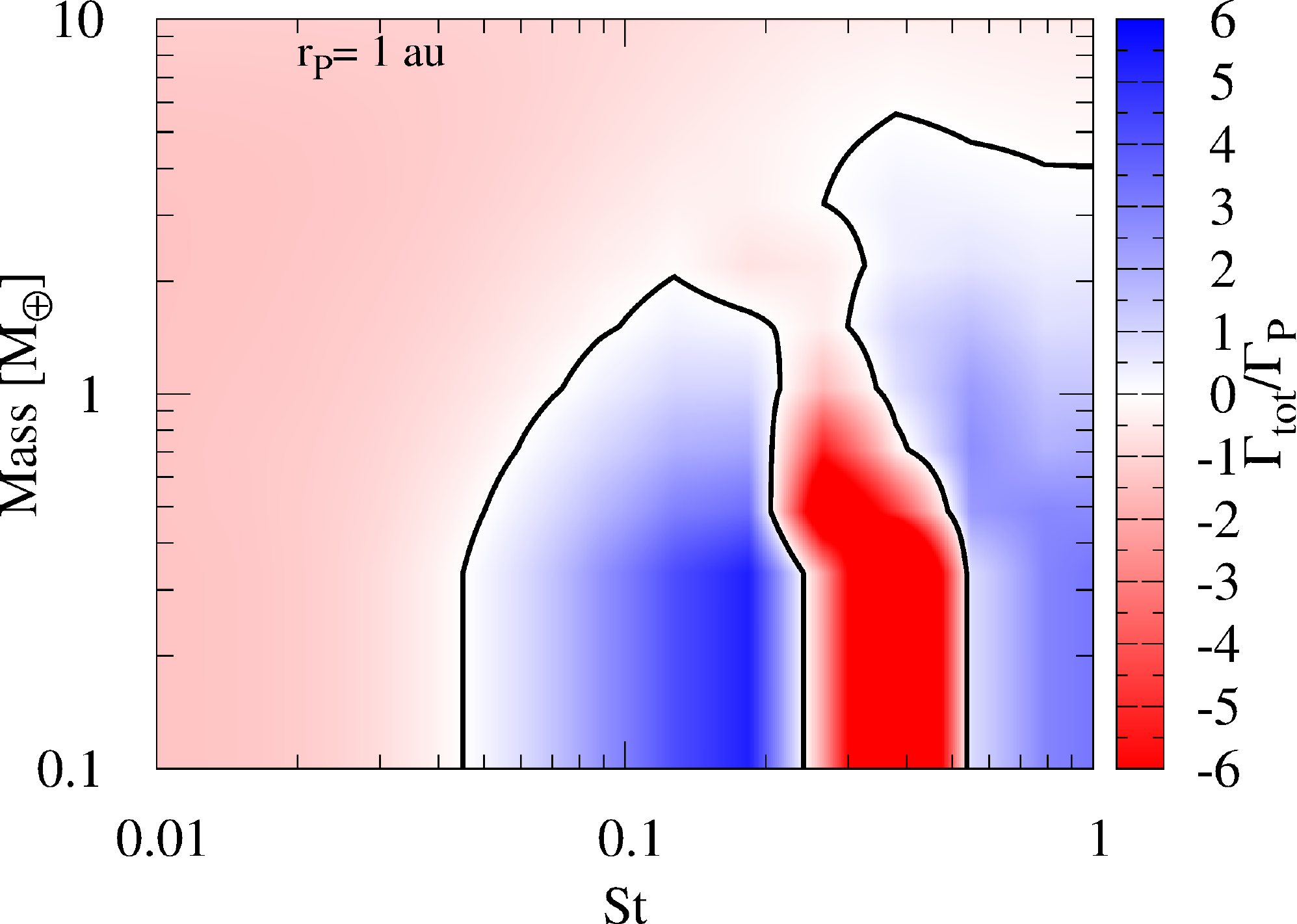} 
     \caption{Same as Fig.~\ref{fig3_sec3}, but considering that the dust torque is obtained using Eq.~(\ref{eq:dust_torque_app}) instead of Eq.~(\ref{eq:dust_torque}).}
    \label{fig2_app1}
\end{figure}

We also note that in Fig.~\ref{fig1_app1} the total torque is lower compared to the total torque in Fig.~\ref{fig:torque_map_mp_vs_rp_non_iso}, where the color scale is strongly saturated (note that $\Gamma_{\text{p}}$ is the same in both cases). This is also appreciable in Fig.~\ref{fig2_app1}, which shows the normalized total torque at 1~au (considering that the dust torque $\Gamma_{\text{d}}$ is given by Eq.~\ref{eq:dust_torque_app}) in a Stokes number -- planet mass diagram as in Fig.~\ref{fig3_sec3}. We remark here that now the gas- and gravity-dominated regimes, where the dust torque generates a total positive torque, appear as disconnected regions in this diagram.    

\begin{figure*}[t!]
    \centering
    \includegraphics[width=0.33\textwidth]{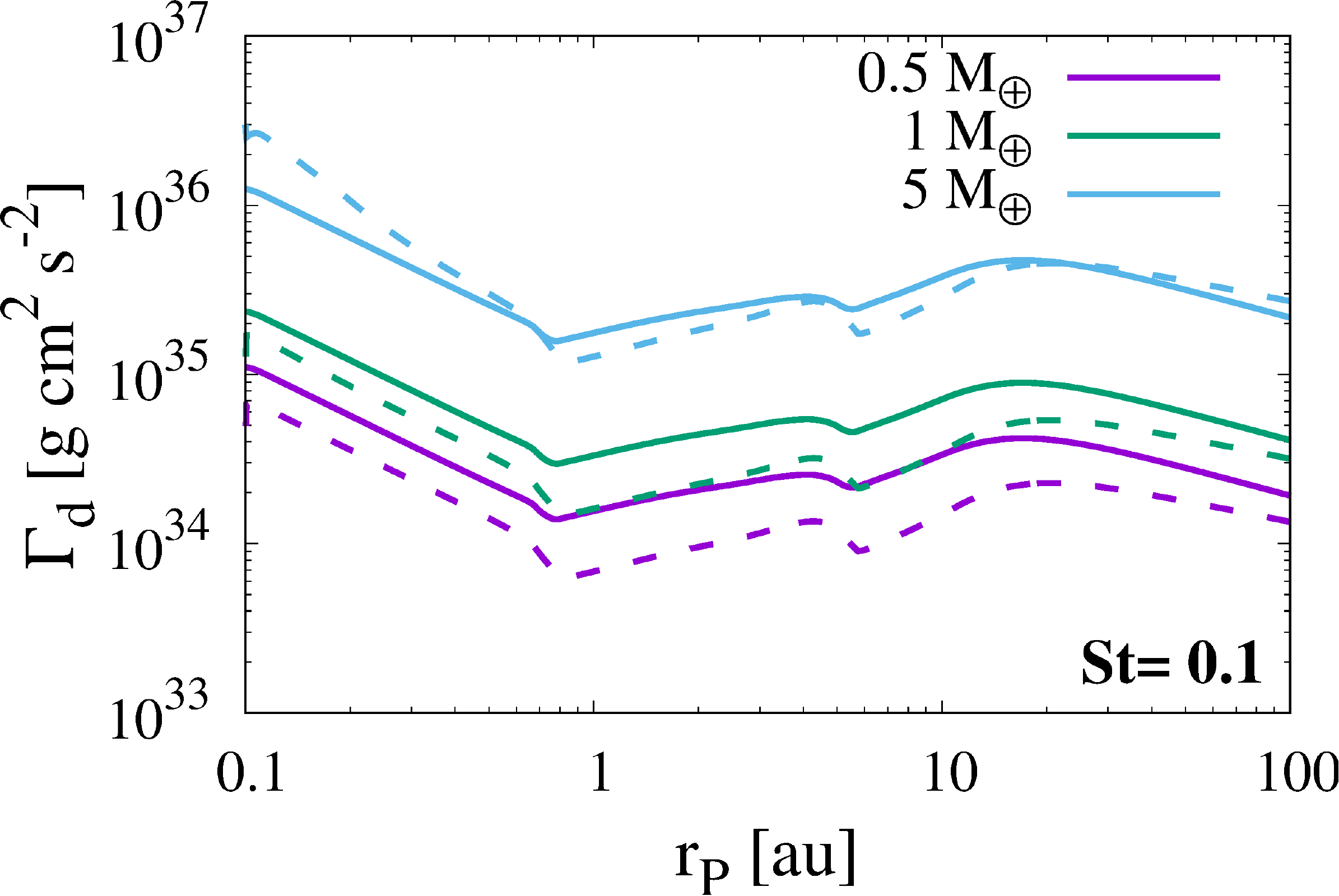}  \includegraphics[width=0.3169\textwidth]{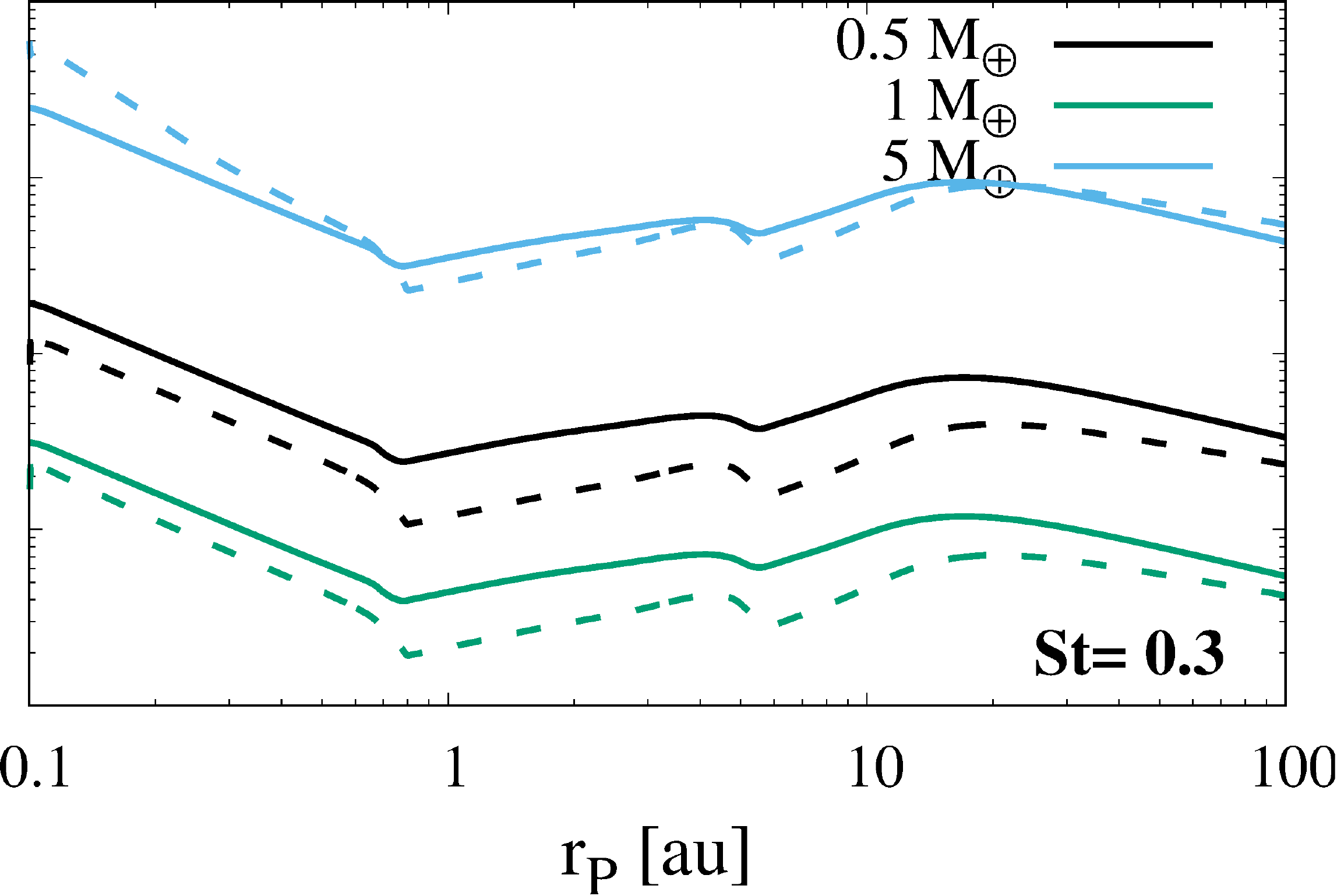} \includegraphics[width=0.3169\textwidth]{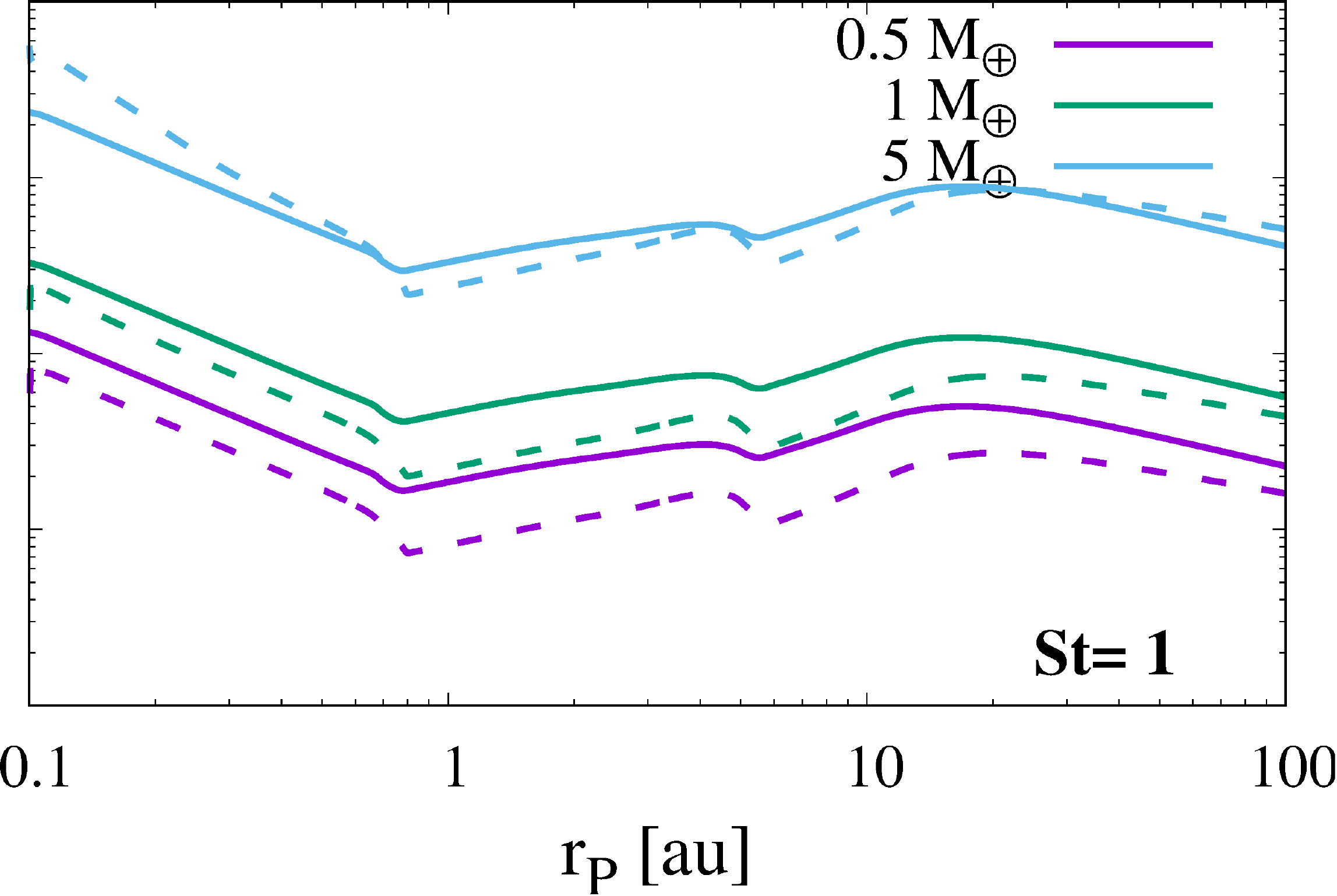}
    \caption{Radial profiles of the dust torque for different planet masses and different Stokes numbers. The solid lines represent the dust torque computed via Eq.~(\ref{eq:dust_torque}), while the dashed lines correspond to the dust torque computed employing Eq.~(\ref{eq:dust_torque_app}). The black lines in the middle panel (the case of St= 0.3) highlight the fact that for a plant mass of $0.5~\text{M}_\oplus$ the dust torques are negative in both approaches.} 
    \label{fig3_app1}
\end{figure*}

In order to assess the impact of the two approaches we outlined to compute the dust torque, we plot in Fig.~\ref{fig3_app1} the radial profile of the dust torque as obtained for three different planet masses, $M_{\rm p} = \{0.5, 1, 5\}\text{M}_\oplus$ and for three different Stokes numbers St= $\{0.1, 0.3, 1\}$ from left to right, respectively. In spite of the fact that the magnitude of the dust torque (shown with solid lines) is in general slightly larger when using Eq.~(\ref{eq:dust_torque}), its radial dependence is qualitative, and even roughly quantitative, similar independently of the approach used to compute it.  

\bibliography{biblio}{}

\begin{thebibliography}{}
\expandafter\ifx\csname natexlab\endcsname\relax\def\natexlab#1{#1}\fi
\providecommand{\url}[1]{\href{#1}{#1}}
\providecommand{\dodoi}[1]{doi:~\href{http://doi.org/#1}{\nolinkurl{#1}}}
\providecommand{\doeprint}[1]{\href{http://ascl.net/#1}{\nolinkurl{http://ascl.net/#1}}}
\providecommand{\doarXiv}[1]{\href{https://arxiv.org/abs/#1}{\nolinkurl{https://arxiv.org/abs/#1}}}

\bibitem[{{Andrews}(2020)}]{Andrews2020}
{Andrews}, S.~M. 2020, \araa, 58, 483,
  \dodoi{10.1146/annurev-astro-031220-010302}

\bibitem[{{Bailli{\'e}} {et~al.}(2016){Bailli{\'e}}, {Charnoz}, \&
  {Pantin}}]{Baillie2016}
{Bailli{\'e}}, K., {Charnoz}, S., \& {Pantin}, E. 2016, \aap, 590, A60,
  \dodoi{10.1051/0004-6361/201528027}

\bibitem[{{Baraffe} {et~al.}(2015){Baraffe}, {Homeier}, {Allard}, \&
  {Chabrier}}]{Baraffe15}
{Baraffe}, I., {Homeier}, D., {Allard}, F., \& {Chabrier}, G. 2015, \aap, 577,
  A42, \dodoi{10.1051/0004-6361/201425481}

\bibitem[{{Ben{\'{\i}}tez-Llambay} {et~al.}(2015){Ben{\'{\i}}tez-Llambay},
  {Masset}, {Koenigsberger}, \& {Szul{\'a}gyi}}]{Benitez-llambay2015}
{Ben{\'{\i}}tez-Llambay}, P., {Masset}, F., {Koenigsberger}, G., \&
  {Szul{\'a}gyi}, J. 2015, \nat, 520, 63, \dodoi{10.1038/nature14277}

\bibitem[{{Ben{\'\i}tez-Llambay} \& {Pessah}(2018)}]{BL2018}
{Ben{\'\i}tez-Llambay}, P., \& {Pessah}, M.~E. 2018, \apjl, 855, L28,
  \dodoi{10.3847/2041-8213/aab2ae}

\bibitem[{{Birnstiel} {et~al.}(2012){Birnstiel}, {Klahr}, \&
  {Ercolano}}]{Birnstiel12}
{Birnstiel}, T., {Klahr}, H., \& {Ercolano}, B. 2012, \aap, 539, A148,
  \dodoi{10.1051/0004-6361/201118136}

\bibitem[{{Bitsch} {et~al.}(2015){Bitsch}, {Johansen}, {Lambrechts}, \&
  {Morbidelli}}]{Bitsch15a}
{Bitsch}, B., {Johansen}, A., {Lambrechts}, M., \& {Morbidelli}, A. 2015, \aap,
  575, A28, \dodoi{10.1051/0004-6361/201424964}

\bibitem[{{Capobianco} {et~al.}(2011){Capobianco}, {Duncan}, \&
  {Levison}}]{Capobianco2011}
{Capobianco}, C.~C., {Duncan}, M., \& {Levison}, H.~F. 2011, \icarus, 211, 819,
  \dodoi{10.1016/j.icarus.2010.09.001}

\bibitem[{{Chambers}(2021)}]{Chambers2021}
{Chambers}, J. 2021, \apj, 914, 102, \dodoi{10.3847/1538-4357/abfaa4}

\bibitem[{{Chen} \& {Lin}(2018)}]{Chen2018}
{Chen}, J.-W., \& {Lin}, M.-K. 2018, \mnras, 478, 2737,
  \dodoi{10.1093/mnras/sty1166}

\bibitem[{{Chrenko} {et~al.}(2017){Chrenko}, {Bro{\v{z}}}, \&
  {Lambrechts}}]{Chrenko2017}
{Chrenko}, O., {Bro{\v{z}}}, M., \& {Lambrechts}, M. 2017, \aap, 606, A114,
  \dodoi{10.1051/0004-6361/201731033}

\bibitem[{{Dittkrist} {et~al.}(2014){Dittkrist}, {Mordasini}, {Klahr},
  {Alibert}, \& {Henning}}]{Dittkrist2014}
{Dittkrist}, K.-M., {Mordasini}, C., {Klahr}, H., {Alibert}, Y., \& {Henning},
  T. 2014, \aap, 567, A121, \dodoi{10.1051/0004-6361/201322506}

\bibitem[{{Dohnanyi}(1969)}]{Dohnanyi69}
{Dohnanyi}, J.~S. 1969, \jgr, 74, 2531, \dodoi{10.1029/JB074i010p02531}

\bibitem[{{Drazkowska} \& {Alibert}(2017)}]{Drazkowska2017}
{Drazkowska}, J., \& {Alibert}, Y. 2017, \aap, 608, A92,
  \dodoi{10.1051/0004-6361/201731491}

\bibitem[{{Drazkowska} {et~al.}(2016){Drazkowska}, {Alibert}, \&
  {Moore}}]{Drazkowska2016}
{Drazkowska}, J., {Alibert}, Y., \& {Moore}, B. 2016, \aap, 594, A105,
  \dodoi{10.1051/0004-6361/201628983}

\bibitem[{{Drazkowska} {et~al.}(2021){Drazkowska}, {Stammler}, \&
  {Birnstiel}}]{Drazkowska21}
{Drazkowska}, J., {Stammler}, S.~M., \& {Birnstiel}, T. 2021, arXiv e-prints,
  arXiv:2101.01728.
\newblock \doarXiv{2101.01728}

\bibitem[{{Guilera} {et~al.}(2023, in prep.){Guilera}, {Benitez-Llambay},
  {Miller Bertolami}, \& {Pessah}}]{Guilera22b}
{Guilera}, O.~M., {Benitez-Llambay}, P., {Miller Bertolami}, M.~M., \&
  {Pessah}, M.~E. 2023, in prep.

\bibitem[{{Guilera} {et~al.}(2019){Guilera}, {Cuello}, {Montesinos}, {Miller
  Bertolami}, {Ronco}, {Cuadra}, \& {Masset}}]{Guilera2019}
{Guilera}, O.~M., {Cuello}, N., {Montesinos}, M., {et~al.} 2019, \mnras, 486,
  5690, \dodoi{10.1093/mnras/stz1158}

\bibitem[{{Guilera} {et~al.}(2021){Guilera}, {Miller Bertolami}, {Masset},
  {Cuadra}, {Venturini}, \& {Ronco}}]{Guilera2021}
{Guilera}, O.~M., {Miller Bertolami}, M.~M., {Masset}, F., {et~al.} 2021,
  \mnras, 507, 3638, \dodoi{10.1093/mnras/stab2371}

\bibitem[{{Guilera} {et~al.}(2017){Guilera}, {Miller Bertolami}, \&
  {Ronco}}]{Guilera2017b}
{Guilera}, O.~M., {Miller Bertolami}, M.~M., \& {Ronco}, M.~P. 2017, \mnras,
  471, L16, \dodoi{10.1093/mnrasl/slx095}

\bibitem[{{Guilera} \& {S{\'a}ndor}(2017)}]{Guilera2017}
{Guilera}, O.~M., \& {S{\'a}ndor}, Z. 2017, \aap, 604, A10,
  \dodoi{10.1051/0004-6361/201629843}

\bibitem[{{Guilera} {et~al.}(2020){Guilera}, {S{\'a}ndor}, {Ronco},
  {Venturini}, \& {Miller Bertolami}}]{Guilera20}
{Guilera}, O.~M., {S{\'a}ndor}, Z., {Ronco}, M.~P., {Venturini}, J., \& {Miller
  Bertolami}, M.~M. 2020, \aap, 642, A140, \dodoi{10.1051/0004-6361/202038458}

\bibitem[{{Hartmann} {et~al.}(1998){Hartmann}, {Calvet}, {Gullbring}, \&
  {D'Alessio}}]{Hartmann98}
{Hartmann}, L., {Calvet}, N., {Gullbring}, E., \& {D'Alessio}, P. 1998, \apj,
  495, 385, \dodoi{10.1086/305277}

\bibitem[{{Jiang} \& {Ormel}(2023)}]{Jiang2023}
{Jiang}, H., \& {Ormel}, C.~W. 2023, \mnras, 518, 3877,
  \dodoi{10.1093/mnras/stac3275}

\bibitem[{{Jim{\'e}nez} \& {Masset}(2017)}]{jm2017}
{Jim{\'e}nez}, M.~A., \& {Masset}, F.~S. 2017, \mnras, 471, 4917,
  \dodoi{10.1093/mnras/stx1946}

\bibitem[{{Kanagawa}(2019)}]{Kanagawa2019}
{Kanagawa}, K.~D. 2019, \apjl, 879, L19, \dodoi{10.3847/2041-8213/ab2a0f}

\bibitem[{{Kunitomo} {et~al.}(2021){Kunitomo}, {Ida}, {Takeuchi}, {Pani{\'c}},
  {Miley}, \& {Suzuki}}]{Kunitomo2021}
{Kunitomo}, M., {Ida}, S., {Takeuchi}, T., {et~al.} 2021, \apj, 909, 109,
  \dodoi{10.3847/1538-4357/abdb2a}

\bibitem[{{Lambrechts} {et~al.}(2014){Lambrechts}, {Johansen}, \&
  {Morbidelli}}]{Lambrechts14}
{Lambrechts}, M., {Johansen}, A., \& {Morbidelli}, A. 2014, \aap, 572, A35,
  \dodoi{10.1051/0004-6361/201423814}

\bibitem[{{Lambrechts} {et~al.}(2019){Lambrechts}, {Morbidelli}, {Jacobson},
  {Johansen}, {Bitsch}, {Izidoro}, \& {Raymond}}]{Lambrechts19}
{Lambrechts}, M., {Morbidelli}, A., {Jacobson}, S.~A., {et~al.} 2019, \aap,
  627, A83, \dodoi{10.1051/0004-6361/201834229}

\bibitem[{{Masset}(2017)}]{masset2017}
{Masset}, F.~S. 2017, \mnras, 472, 4204, \dodoi{10.1093/mnras/stx2271}

\bibitem[{{McNally} {et~al.}(2019){McNally}, {Nelson}, {Paardekooper}, \&
  {Ben{\'\i}tez-Llambay}}]{McNally2019}
{McNally}, C.~P., {Nelson}, R.~P., {Paardekooper}, S.-J., \&
  {Ben{\'\i}tez-Llambay}, P. 2019, \mnras, 484, 728,
  \dodoi{10.1093/mnras/stz023}

\bibitem[{{Miguel} {et~al.}(2011){Miguel}, {Guilera}, \&
  {Brunini}}]{Miguel2011}
{Miguel}, Y., {Guilera}, O.~M., \& {Brunini}, A. 2011, \mnras, 412, 2113,
  \dodoi{10.1111/j.1365-2966.2010.17887.x}

\bibitem[{{Miranda} \& {Rafikov}(2019)}]{Miranda2019}
{Miranda}, R., \& {Rafikov}, R.~R. 2019, \apjl, 878, L9,
  \dodoi{10.3847/2041-8213/ab22a7}

\bibitem[{{Morbidelli}(2020)}]{Morby2020}
{Morbidelli}, A. 2020, arXiv e-prints, arXiv:2004.04942.
\newblock \doarXiv{2004.04942}

\bibitem[{{Morbidelli} \& {Raymond}(2016)}]{2016JGRE..121.1962M}
{Morbidelli}, A., \& {Raymond}, S.~N. 2016, Journal of Geophysical Research
  (Planets), 121, 1962, \dodoi{10.1002/2016JE005088}

\bibitem[{{Ogihara} \& {Hori}(2020)}]{Ogihara2020}
{Ogihara}, M., \& {Hori}, Y. 2020, arXiv e-prints, arXiv:2003.05934.
\newblock \doarXiv{2003.05934}

\bibitem[{{Paardekooper}(2014)}]{Paardekooper2014}
{Paardekooper}, S.~J. 2014, \mnras, 444, 2031, \dodoi{10.1093/mnras/stu1542}

\bibitem[{{Paardekooper} {et~al.}(2022){Paardekooper}, {Dong}, {Duffell},
  {Fung}, {Masset}, {Ogilvie}, \& {Tanaka}}]{Paardekooper2022}
{Paardekooper}, S.-J., {Dong}, R., {Duffell}, P., {et~al.} 2022, arXiv
  e-prints, arXiv:2203.09595.
\newblock \doarXiv{2203.09595}

\bibitem[{{Pierens} {et~al.}(2019){Pierens}, {Lin}, \& {Raymond}}]{Pierens2019}
{Pierens}, A., {Lin}, M.~K., \& {Raymond}, S.~N. 2019, \mnras, 488, 645,
  \dodoi{10.1093/mnras/stz1718}

\bibitem[{{Pinilla} {et~al.}(2012){Pinilla}, {Birnstiel}, {Ricci}, {Dullemond},
  {Uribe}, {Testi}, \& {Natta}}]{Pinilla2012}
{Pinilla}, P., {Birnstiel}, T., {Ricci}, L., {et~al.} 2012, \aap, 538, A114,
  \dodoi{10.1051/0004-6361/201118204}

\bibitem[{{Reg{\'a}ly}(2020)}]{Regaly2020}
{Reg{\'a}ly}, Z. 2020, \mnras, 497, 5540, \dodoi{10.1093/mnras/staa2181}

\bibitem[{{Ronco} {et~al.}(2017){Ronco}, {Guilera}, \& {de
  El{\'\i}a}}]{Ronco2017}
{Ronco}, M.~P., {Guilera}, O.~M., \& {de El{\'\i}a}, G.~C. 2017, \mnras, 471,
  2753, \dodoi{10.1093/mnras/stx1746}

\bibitem[{{Schneider} \& {Bitsch}(2021)}]{Schneider21}
{Schneider}, A.~D., \& {Bitsch}, B. 2021, \aap, 654, A71,
  \dodoi{10.1051/0004-6361/202039640}

\bibitem[{{Shakura} \& {Sunyaev}(1973)}]{SS73}
{Shakura}, N.~I., \& {Sunyaev}, R.~A. 1973, \aap, 24, 337

\bibitem[{{Stammler} \& {Birnstiel}(2022)}]{Stammler2022}
{Stammler}, S.~M., \& {Birnstiel}, T. 2022, \apj, 935, 35,
  \dodoi{10.3847/1538-4357/ac7d58}

\bibitem[{{Tanaka} {et~al.}(2002){Tanaka}, {Takeuchi}, \& {Ward}}]{Tanaka02}
{Tanaka}, H., {Takeuchi}, T., \& {Ward}, W.~R. 2002, \apj, 565, 1257,
  \dodoi{10.1086/324713}

\bibitem[{{Venturini} {et~al.}(2020{\natexlab{a}}){Venturini}, {Guilera},
  {Haldemann}, {Ronco}, \& {Mordasini}}]{Venturini20Letter}
{Venturini}, J., {Guilera}, O.~M., {Haldemann}, J., {Ronco}, M.~P., \&
  {Mordasini}, C. 2020{\natexlab{a}}, \aap, 643, L1,
  \dodoi{10.1051/0004-6361/202039141}

\bibitem[{{Venturini} {et~al.}(2020{\natexlab{b}}){Venturini}, {Guilera},
  {Ronco}, \& {Mordasini}}]{Venturini20ST}
{Venturini}, J., {Guilera}, O.~M., {Ronco}, M.~P., \& {Mordasini}, C.
  2020{\natexlab{b}}, \aap, 644, A174, \dodoi{10.1051/0004-6361/202039140}

\bibitem[{{Venturini} {et~al.}(2020{\natexlab{c}}){Venturini}, {Ronco}, \&
  {Guilera}}]{Venturini20Review}
{Venturini}, J., {Ronco}, M.~P., \& {Guilera}, O.~M. 2020{\natexlab{c}}, \ssr,
  216, 86, \dodoi{10.1007/s11214-020-00700-y}

\bibitem[{{Voelkel} {et~al.}(2022){Voelkel}, {Klahr}, {Mordasini}, \&
  {Emsenhuber}}]{Voelkel2022}
{Voelkel}, O., {Klahr}, H., {Mordasini}, C., \& {Emsenhuber}, A. 2022, arXiv
  e-prints, arXiv:2202.01500.
\newblock \doarXiv{2202.01500}

\bibitem[{{Weber} {et~al.}(2018){Weber}, {Ben{\'\i}tez-Llambay}, {Gressel},
  {Krapp}, \& {Pessah}}]{Weber2018}
{Weber}, P., {Ben{\'\i}tez-Llambay}, P., {Gressel}, O., {Krapp}, L., \&
  {Pessah}, M.~E. 2018, \apj, 854, 153, \dodoi{10.3847/1538-4357/aaab63}

\bibitem[{{Zhu} {et~al.}(2012){Zhu}, {Nelson}, {Dong}, {Espaillat}, \&
  {Hartmann}}]{Zhu2012}
{Zhu}, Z., {Nelson}, R.~P., {Dong}, R., {Espaillat}, C., \& {Hartmann}, L.
  2012, \apj, 755, 6, \dodoi{10.1088/0004-637X/755/1/6}

\end{thebibliography}
\bibliographystyle{aasjournal}



\end{document}